\documentclass[twocolumn,superscriptaddress,prd,showkeys,showpacs,nofootinbib]{revtex4-2}

\usepackage[colorlinks = true,
            linkcolor = cyan,
            urlcolor  = blue,
            citecolor = red,
            anchorcolor = blue]{hyperref}\usepackage{color}
\usepackage{amssymb}
\usepackage[nointegrals]{wasysym}
\usepackage{amsthm}
\usepackage{textcomp}
\usepackage{mathtools}
\usepackage{comment}
\usepackage{graphicx}
\usepackage{soul}
\usepackage{multirow}
\usepackage{ulem}

\usepackage{lineno}

\usepackage{comment}
\usepackage[separate-uncertainty,retain-explicit-plus,per-mode = symbol]{siunitx}

\usepackage{url}

\newcommand{\iso}[2]{\mbox{$^{#1}$#2}}
\newcommand{\us}{\mbox{$\mu$s}}

\begin{document}

\title{Gas Electroluminescence in a Dual Phase Xenon-Doped Argon Detector}

\author{J.W.~Kingston} \email{kingston2@llnl.gov}\affiliation{Lawrence Livermore National Laboratory, 7000 East Ave., Livermore, CA 94550, USA}
\affiliation{University of California Davis, Department of Physics, One Shields Ave., Davis, CA 95616, USA}
\author{J.~Qi} \email{qi5@llnl.gov} \affiliation{Lawrence Livermore National Laboratory, 7000 East Ave., Livermore, CA 94550, USA}
\affiliation{University of California San Diego, Department of Physics, La Jolla, CA 92093-0112, USA}
\author{J.~Xu} \affiliation{Lawrence Livermore National Laboratory, 7000 East Ave., Livermore, CA 94550, USA}
\author{E.P.~Bernard} \affiliation{Lawrence Livermore National Laboratory, 7000 East Ave., Livermore, CA 94550, USA}
\author{A.D. Tidball}
\affiliation{University of California Davis, Department of Physics, One Shields Ave., Davis, CA 95616, USA}
\author{A.W. Peck}
\affiliation{University of California Riverside, Physics and Astronomy Department, 900 University Ave., Riverside, CA 92521, USA}
\author{N.S.~Bowden} \affiliation{Lawrence Livermore National Laboratory, 7000 East Ave., Livermore, CA 94550, USA}
\author{M. Tripathi}
\affiliation{University of California Davis, Department of Physics, One Shields Ave., Davis, CA 95616, USA}
\author{K.~Ni} \affiliation{University of California San Diego, Department of Physics, La Jolla, CA 92093-0112, USA}
\author{S. Westerdale}
\affiliation{University of California Riverside, Physics and Astronomy Department, 900 University Ave., Riverside, CA 92521, USA}

\date{\today}

\begin{abstract}
Noble element detectors using argon or xenon as the detection medium are widely used in the searches for rare neutrino and dark matter interactions. Xenon doping in liquid argon can preserve attractive properties of an argon target while enhancing the detectable signals with properties of xenon. 
In this work, we deployed a dual-phase liquid argon detector with up to 4\% xenon doping in the liquid and studied its gas electroluminescence properties as a function of xenon concentration. At $\sim$2\% xenon doping in liquid argon, we measured $\sim$34 ppm of xenon in the gas and observed $\sim$2.5 times larger electroluminescence signals using VUV silicon photomultipliers than those in pure argon. Analysis of signals of different wavelengths  confirms that the argon gas electroluminescence process is strongly modified by the addition of xenon. 
We propose an analytical model to describe the underlying energy transfer mechanism in argon-xenon gas mixtures. Lastly, the implications of this measurement for low energy ionization signal detection will be discussed. 
\end{abstract}


\maketitle

%\linenumbers

\section{Introduction}
\label{sec:introdcution}

Liquid argon ionization detectors are widely used in rare event searches~\cite{ICARUS1997_ArTPC, Bonivento2024_ArTPC,DarkSide2018_532day}.
Their sensitivity to rare processes can partially be owed to their scalable nature \cite{AbedAbud_2022, agostino2014lbnodemolargescaleneutrinodetector}, as well as their ability to be made chemically and radioactively pure \cite{Darkside20kCleanAr}. Energy deposited from particle interactions in liquid argon is released in three avenues: scintillation photons, ionization electrons, and heat. Scintillation photons can be detected using photosensors such as photomultiplier tubes (PMTs) or silicon photomultipliers (SiPMs). In the presence of an electric field,
a fraction of ionization electrons can escape recombination with the ionized argon atoms and drift through the liquid bulk towards a charge detection region, which could be a charge readout plane in a single phase detector \cite{Acciarri_2017}, or a gas electroluminescence (EL) setup in a dual phase detector \cite{Akerib_2020}. Heat is usually not detected. 

A dual-phase argon time projection chamber (TPC) gains improved sensitivity to low-energy ionization signals from the amplification of ionization signals through gas EL. Each single electron extracted into the gas region typically produces hundreds of secondary photons~\cite{Monteiro2008_ArEL,Oliveira2011_EL}, enabling a TPC to be sensitive to single electron signals~\cite{DS50_2018_S2Only}. 
Therefore, a liquid argon TPC is a promising choice for detecting dark matter interactions or Coherent Elastic Neutrino-Nucleus Scattering (CEvNS), both of which are expected to produce mostly low energy signals~\cite{DarkSide2018_532day,Essig2022_LowMassDM}. 
In addition, the lower atomic mass of argon relative to xenon enables it to receive more kinetic energy from a low-mass dark matter particle or neutrino, and the lower temperature of liquid argon also suppresses the outgassing of impurities from detector materials. 
As demonstrated in both liquid argon and xenon, impurities in the liquid can capture drift electrons and release them at a later time as a background~\cite{LUX2020_SE,50collaboration2025characterizationspuriouselectronsignalsdoublephase,xenon1t_ebg_2021}. 
While excessive electron background has been observed in both argon and xenon TPCs, the background level appears lower in liquid argon \cite{Akerib_2020_ee,50collaboration2025characterizationspuriouselectronsignalsdoublephase}. 

In practice, the detection of low-energy ionization signals in liquid argon is subject to unique difficulties. 
At the time of writing, the ionization signals measured in argon TPCs have not enabled a position and energy reconstruction accuracy comparable to those achieved in xenon~\cite{LUX2018_PositionReconstruction,LZ2023_EnergyReconstruction}. 
This may be partially attributed to the difficulty in detecting the short-wavelength (128 nm) light from argon scintillation or EL. In current argon detectors, wavelength shifting (WLS) chemicals such as Tetraphenyl Butadiene (TPB) are coated onto detector or photosensor surfaces to shift the 128 nm photons into the visible range before they can be detected~\cite{DS10_2013,DS502015_Detector}. 
Despite high light yield observed in such configurations, the intermediate photon absorption and re-emission process in the WLS causes some information, such as the original photon emission position, to be lost. 
In addition, spatial and temporal variations of the WLS coating quality as well as the stochastic nature of its fluorescence can further complicate the light collection process.

Direct detection of 128 nm argon light has been made possible by new vacuum ultraviolet (VUV) photosensors, such as PMTs with VUV-transmissive windows or VUV SiPMs. 
However, these devices typically have a much lower quantum efficiency (QE) for 128 nm light than for longer wavelength photons such as those from xenon scintillation (175 nm) \cite{HamamatsuVUVGraph}. 
In addition, few reflectors can efficiently reflect 128 nm photons to facilitate light collection. 
Combining these practical constraints with the higher excitation energy of argon atoms for EL than xenon, even a liquid argon TPC instrumented with VUV SiPMs or VUV PMTs may not match the ionization detection performance of a xenon TPC. 

In reference \cite{CHILLAX2023_Thermodynamics}, we proposed that the challenge of argon EL detection can be addressed with xenon doping in liquid argon, which also brings additional benefits. 
It is known that a small amount of xenon in liquid argon ($\sim$10 ppm) can efficiently transfer energy from argon excitation to xenon, leading to the emission of scintillation light in the xenon wavelength (175 nm)~\cite{Neumeier_2015, DUNE:2024dge}. 
A similar transfer of argon excitation energy to xenon can be expected for gas EL; we calculated that a few percent of xenon doping in liquid argon is required to produce tens of ppm of xenon concentration in the gas and to enable this process. 
Production of xenon EL light will improve the EL signal detection with commercial photosensors and common reflectors such as PTFE. 
In addition, the substantial amount of xenon present in liquid argon could also enhance the ionization yield of both nuclear recoils and electron recoils~\cite{Kubota1976_ChargeYield} through direct ionization or Penning ionization, thanks to the low ionization energy of xenon. 
Lastly, although the xenon concentration in the gas phase is only in the ppm range, its low excitation energy could lead to direct excitation of xenon atoms by accelerated electrons and enhance the EL yield. 

In this work, we deploy a compact, dual-phase, argon TPC with percent-level xenon doping to study the gas EL properties at different xenon concentrations, electric field strengths and gas pressures. 
This work builds on our ability to reliably dope large amounts of xenon into liquid argon and to maintain relatively stable detector operation. 
In Sec.~\ref{sec:apparatus} we describe the experimental apparatus and the type of data we take; 
in Sec.~\ref{sec:result} we analyze the measured data and quantify the increase in detected EL signals; Section~\ref{sec:s2_shape} presents an analytical model of the underlying physics related to EL in argon-xenon mixtures;
in Sec.~\ref{sec:disc} we discuss the major results and their implications for future ionization signal detection in liquid argon, and 
in Sec.~\ref{sec:concl} we conclude the work. 

\section{Experimental Setup}
\label{sec:apparatus}

The measurements presented in this work were carried out with the CHILLAX (CoHerent Ionization Limit in Liquid Argon and Xenon) setup. 
The cryogenic system of CHILLAX was designed to stabilize percent-level xenon-doped liquid argon. As discussed in our prior work~\cite{CHILLAX2023_Thermodynamics}, we demonstrated stability of 2.35\% xenon in liquid argon (by molar fraction) for several days. We have since refined both detector control and operating techniques, leading to an improved stabilization capability that can handle up to 5\% doping indefinitely. The core cryogenic system remains largely unchanged and details can be found in \cite{CHILLAX2023_Thermodynamics}.

\begin{figure}[t!]
    \centering
    \includegraphics[width=0.99\columnwidth]{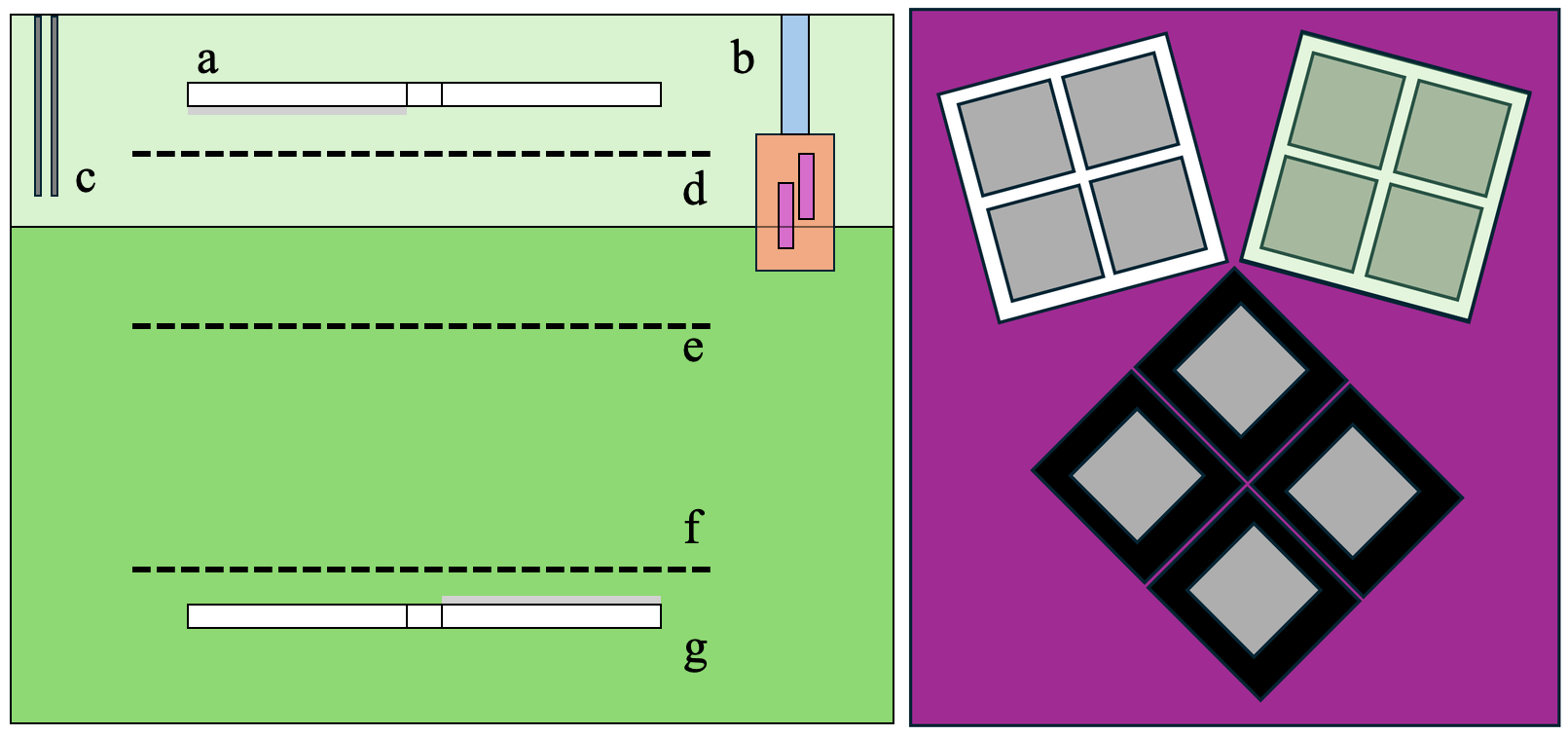}
    \caption{\textbf{Left:} A schematic of the CHILLAX TPC (not to scale). The top SiPM assembly (a) sits in the gas phase (light green) and looks through the anode (d) into the liquid bath (dark green), where the cathode (e) defines the lower boundary of the active volume. Below the cathode lies a grounded shield (f) that screens the bottom SiPM assembly (g). The liquid level height is set using a two-RTD level meter (b) and the xenon concentration in the gas phase is sampled through a capillary tube (c). \textbf{Right:} A to-scale schematic of the top SiPM assembly. The upper right SiPM is a S13371 model masked with a quartz window, and the upper left SiPM is the same model with the window removed. The lower SiPM is a 2x2 array of windowless S13370 units, but is read out as a single channel in the same way as the S13371s.}
    \label{fig:chillaxSchematic}
\end{figure}

For this gas EL study, we installed a compact dual-phase TPC in CHILLAX, a schematic for which is shown in Figure \ref{fig:chillaxSchematic}. 
The active target is a \diameter 5.7 cm $\times$ 0.96 cm cylindrical column that is filled with pure liquid argon or a liquid argon-xenon mixture. 
The target is bound from below by a cathode grid and above by a gas layer. An anode grid is suspended 0.76 cm above the liquid surface, defining the EL gas gap. Each grid holder ring has an OD of 6.4 cm and an ID of 5.7 cm, supporting an etched hexagonal mesh grid with 89\% transparency at normal incidence. 
The liquid surface is set between two PT100 Resistance Temperature Detectors (RTDs) offset in height by 1.2 mm. Despite the nearly identical temperatures of the liquid and the gas near the liquid surface, a powered RTD contacting the liquid measures a lower resistance value than one surrounded by gas due to the higher thermal conductivity of the liquid relative to the gas. 
This gap contributes a $\pm$0.6 mm uncertainty to the heights of the active TPC target and the gas EL region. 

During operations, the cathode is maintained at a negative high voltage (HV) to provide an electric field, wherein ionization electrons produced in the target volume are drifted through the liquid bulk and extracted into the gas phase to generate excimer EL light.
The EL light is then detected by an assembly comprised of 3 Hamamatsu VUV4-series SiPM modules above the anode, and a matching assembly below the cathode. 
Each assembly consists of a 2x2 array of S13370-6075CN units (6x6 $\text{mm}^2$ each), which are grouped together for signal readout, and two individual S13371-6050CQ-02 units, which each contains a similar 2x2 array in a more compact layout. A schematic of the SiPM layout is illustrated in Fig.~\ref{fig:chillaxSchematic} (right). 
One of the two S13371-6050CQ-02 modules has a quartz window in front of the sensitive SiPM area, absorbing photons with wavelengths below 160 nm; the other S13371-6050CQ-02 module has its quartz window removed~\cite{Pershing_2022} to be sensitive to short wavelength photons, including 128 nm (argon excimer) and an intermediate  wavelength of $\sim$147 nm reported in gaseous argon and xenon mixtures~\cite{ Efthimiopoulos1997,Takahashi1975_Xe}. 
The windowless S13371, the windowed S13371 and the 2x2 S13370 in the top assembly are labeled as channel 0, 1, and 2, respectively, with their counterparts in the bottom assembly designated as channel 3, 4 and 5.
The bottom SiPM assembly is shielded from the cathode HV by an additional grid at ground potential, and the top assembly is shielded by the grounded anode. 
For the measurements, the S13370s are biased at 1.8 V over-voltage and the S13371s are biased at 4.4 V over-voltage to approximately match the SPE gain between models. 
The output of every SiPM channel passes through a 23 MHz two-pole LC low-pass filter and then a $\times$10 amplifier.

The detector starts with pure liquid argon in the active region, and then the xenon concentration in the liquid is increased stepwise to 1.17\%, 2.07\%, 3.03\%, and 3.81\% by molar fraction. 
For convenience, we hereafter refer to these xenon concentrations as 1\%, 2\%, 3\% and 4\% [Xe], respectively. 
For most of the data taking, the system absolute pressure is maintained at 2.125 $\pm$ 0.002 bar, corresponding to a temperature of 95.0 $\pm$ 0.5 K in pure argon. 
With every 1\% of xenon added, the temperature of the mixtures climbs by approximately 0.1 K to maintain the constant system pressure due to the low vapor pressure of xenon that replaces argon. 
This observation is consistent with Raoult's law and measurements in the literature. 
Every time xenon is added to the system, the liquid level slightly increases, and is reset to the target level by venting argon gas in controlled quantities. This loss in argon mass is taken into account when the xenon concentration in the liquid is calculated.
Nevertheless, the TPC target mass still increases due to changes in liquid density: at the aforementioned xenon concentrations, the target mass is calculated to be 32.64 g, 33.81 g, 34.65 g, 35.50 g, and 36.17 g, respectively. 

Gas from the detector is continuously circulated in a closed loop through a pump and a SAES MonoTorr getter for gas-phase purification at a rate of 1 Standard Liter Per Minute (SLPM) during steady state operation. 
Impurity species more volatile than argon are efficiently removed. 
Purified gas is re-condensed in a vessel cooled with a heat exchanger (HX), and then flows into the main detector bath. 
This liquid flow from the HX to the detector also entrains xenon that is injected with the circulated gas into the HX during doping and brings it into the main detector volume.
The xenon concentration in the gas above the detector liquid is suppressed by a factor of $\sim$625 relative to that in the liquid according to Henry's Law. 
After each doping step, we draw gas samples from near the liquid surface though a purgeable stainless capillary (0.07 cm ID) extending from the top of the cryostat down to approximately 1 cm above the gas-liquid interface. At 1\%, 2\%, 3\%, and 4\% [Xe] in the liquid, the  xenon concentrations in the gas phase were measured to be $14\pm4$~ppm, $34\pm4$~ppm, $57\pm4$ppm, and $61\pm6$~ppm, in approximate agreement with the predicted values of 19 ppm, 32 ppm, 49 ppm, and 61 ppm using  Henry's Law. 
Because the circulated gas is mostly argon, its condensation in the HX lowers the xenon concentration in the HX liquid toward that in the circulated gas after xenon doping is stopped, and essentially all doped xenon is delivered to the detector bath. 

During the measurement campaign, the TPC was operated with cathode high voltage values of -10 kV, -12 kV, and -14 kV, which henceforth will be referenced by absolute value. 
These voltages correspond to electric fields of 4.6 $\pm$ 0.1 kV/cm, 5.5 $\pm$ 0.2 kV/cm, and 6.4 $\pm$ 0.2 kV/cm respectively in the liquid and 6.8 $\pm$ 0.1 kV/cm, 8.2 $\pm$ 0.2 kV/cm, and 9.6 $\pm$ 0.2 kV/cm respectively in the gas. 
The electric field values are computed by COMSOL electric field simulations, using the detector geometry imported from SolidWorks, and averaged over a cylindrical fiducial volume, which is radially bounded by the SiPMs and excludes 1 mm from the cathode and anode grids.
These values deviate from a parallel plate approximation by -3.0\% in the liquid and -4\% in the gas. The main sources of uncertainty are the spatial variation of the field in the fiducial volume and the variation of the liquid level within the bounds of the liquid level meter. The change in liquid dielectric constant is small from added xenon ($<$1\% at 4\% [Xe]), so we approximate a constant electric field across all xenon levels.
At the operating pressure of 2.125 bar, these values correspond to reduced electric fields in the gas of 4.0 $\pm$ 0.1 Td, 4.8 $\pm$ 0.1 Td, and 5.6 $\pm$ 0.1 Td, respectively.

For each combination of xenon concentration and electric field, we produce gas EL signals by irradiating the active liquid volume with low-energy gamma rays. 
A 17 MBq \iso{241}{Am} check source is mounted on the outer wall of the vacuum insulation chamber and above the liquid surface to minimize attenuation in inactive liquid volumes around the TPC and to create a relatively uniform gamma illumination in the active liquid. 
\iso{241}{Am} primarily alpha decays to \iso{237}{Np} with the emission of a 59.5 keV gamma ray at a branching ratio of 36\%, 
providing a quasi-mono-energetic signal in the detector.
The observed \iso{241}{Am} event rate was 80 Hz - 250 Hz in the TPC, and we typically record a total of 300,000 events per configuration. 
For each detector configurations we also acquired background data with no source, with a typical event rate of 2 Hz --10 Hz; usually 50,000 events were acquired. 
In a small number of acquisitions fewer events were recorded due to HV instability issues. 

The acquired data consists of individual waveforms from 6 SiPM channels digitized with a Struck SIS3316 digitizer at a 250 MHz speed and with 14 bit precision. 
The grouped SiPM readout scheme in our setup increases the device capacitance, leading to a typical SPE rise time of 50 ns; this allows us to down-sample the recorded waveforms in analysis by a factor of 16 for faster processing after smoothing them with a 200 ns boxcar filter. 
The digitizer is triggered when the amplitudes of two or more channel waveforms exceed a preset threshold relative to the baseline within a 20 ns coincidence window. Once triggered, the digitizer records 20 \us\ of waveform before the trigger time and 80 \us\ after the trigger time for all 6 channels. Henceforth, a single instance of the described trigger and waveform acquisition in all channels is referred to as a single ``event''
. The same trigger setting and acquisition window was utilized across all datasets.

\section{Electroluminscence Gain}
\label{sec:result}

For each SiPM waveform, an adaptive baseline algorithm dynamically adjusts the DC offset in the waveform to capture low-frequency fluctuations in the baseline, which is then subtracted to allow for pulses as small as single photoelectrons (SPEs) to be efficiently identified. 
The SPE size of individual SiPM channels in each dataset is determined by sampling isolated spikes in the tail of their respective waveforms and fitting their integral distribution to a Gaussian. 
Because of the small SPE variation between datasets ($\textless$ 2.2\%), the average SPE value over all datasets for each channel is used in this work. The integrals of S2 pulses are then converted from units of ADC counts $\times$ ns to photoelectrons (PEs).

The baseline-subtracted and SPE-calibrated waveforms from all 6 SiPMs are also summed over to produce event-level waveforms. 
Pulse reduced-quantities (RQs) such as the time, amplitude, integral, and pulse shape parameters are computed for further analysis.

\subsection{Event Selection}
\label{sec:eventselection}

\begin{figure}[!htbp]
    \centering
    \includegraphics[width=0.95\columnwidth]{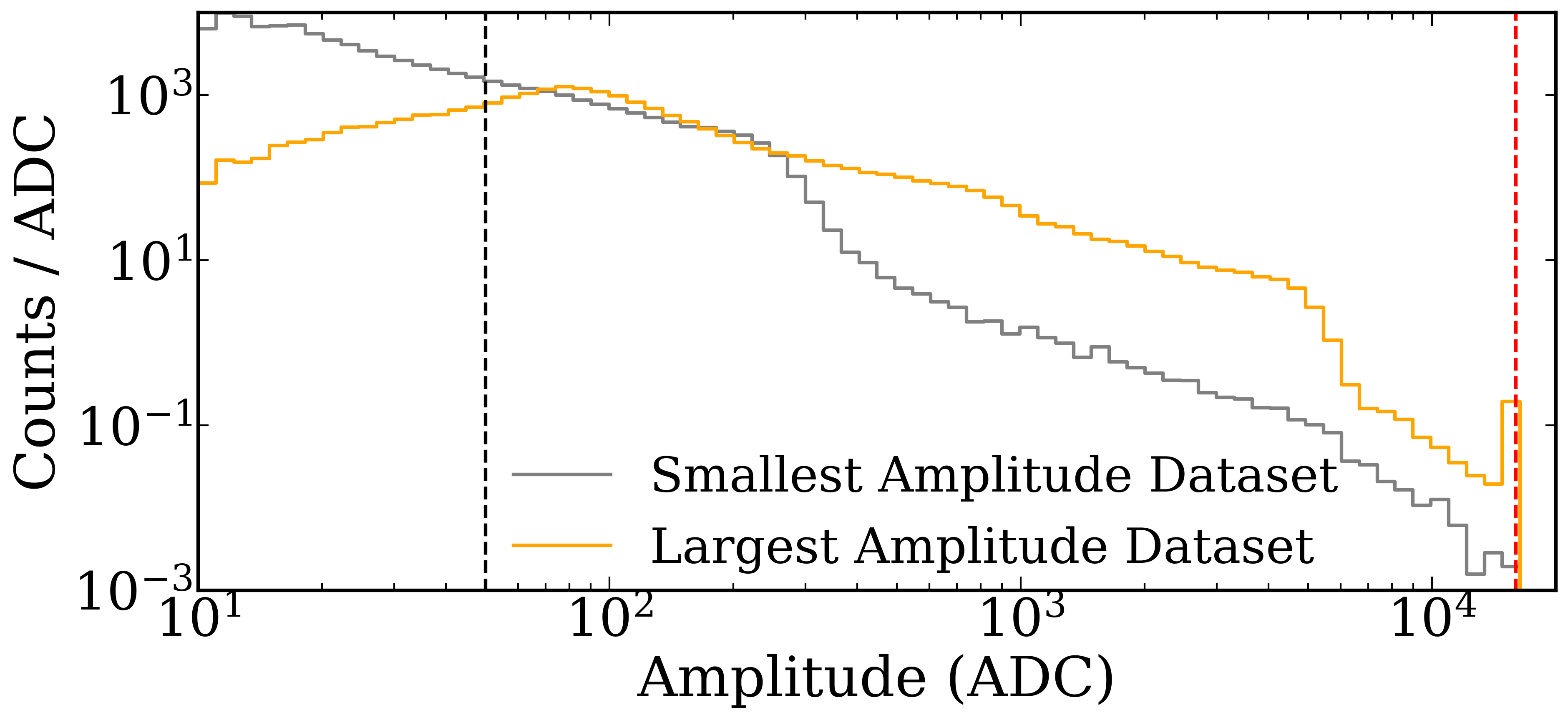}
    \caption{Pulse amplitude distributions corresponding to the smallest amplitude dataset (gray, 10 kV, 0\% [Xe], Ch 2) and the largest (orange, 14 kV, 4\% [Xe], Ch 0) superimposed with the saturation and trigger threshold (red and black, respectively). The shoulders at 300 ADC and 6000 ADC in the gray and orange spectra respectively correspond to the \iso{241}{Am} events with the largest amplitude. The counts with amplitudes larger than these shoulders correspond to background events.}
    \label{fig:ampChecks}
\end{figure}

An implicit event selection criterion is the data acquisition trigger. The trigger for every dataset was set to 2-fold coincidence among 5 out of the 6 SiPMs (1 channel exhibited high bipolar noise rates and was excluded) with a threshold of 50 ADC counts ($\sim$6 mV).
Figure~\ref{fig:ampChecks} shows the pulse amplitude distributions for \iso{241}{Am} events with the lowest (Ch 2, 10 kV cathode voltage, 0\% [Xe]) and highest (Ch 0, 14 kV cathode voltage, 4\% [Xe]) gain settings. 
The ``shoulder''-like feature around 300 ADC and 6,000 ADC in the two spectra correspond to the largest signals induced by \iso{241}{Am} energy depositions in the liquid and are henceforth referred to as the endpoints. 
Even at the lowest gain setting, the pulse amplitudes for the endpoint events are substantially above the trigger threshold. 
Similarly, we chose a detector configuration to avoid saturating the 2V dynamic input range of the digitzer and other signal processing electronics for the dataset with the largest S2 gain, which has a maximum endpoint pulse amplitude of $\sim$7000 ADC, approximately midrange for our 14-bit digitizer ($2^{14} = 16384$ ADC). 
Saturation effects are observed for high-energy background events in datasets with large gains, as shown in Fig.~\ref{fig:ampChecks}, however, our analysis focuses primarily on events near the \iso{241}{Am} endpoint. Therefore, we do not anticipate significant bias from trigger threshold or digitizer saturation for the events of interest.

A dual-phase noble liquid TPC collects both scintillation and ionization signals generated at the energy deposition site, which are colloquially referred to as S1 and S2 signals, respectively. Because of the EL gain for an ionization signal, the detected S2 pulse is usually much larger than the associated S1. 
The time separation between the S1 and S2 pulses is determined by the depth of the event under the liquid surface and the electron drift speed in the liquid. 
For our detector geometry and electric fields, the electron drift time through the active region can range from 0 to 3 $\mu$s \cite{Gushchin1980Drifts}. As a result, a significant fraction of the S1 pulses is identified within the same pulse window as S2s. 
Events with two or more large pulses identified are mostly from multi-scatter events and erroneous pulse finding results, so an event is discarded if the second-largest pulse area is greater than 10\% of the largest one. 
We also reject events that contain a pulse in the first 2 \us\ of the event window, a trait typical of events triggered by continuous light emission produced by high voltage instabilities. 

\begin{figure}[t!]
  \centering
    \includegraphics[width=0.8\columnwidth]{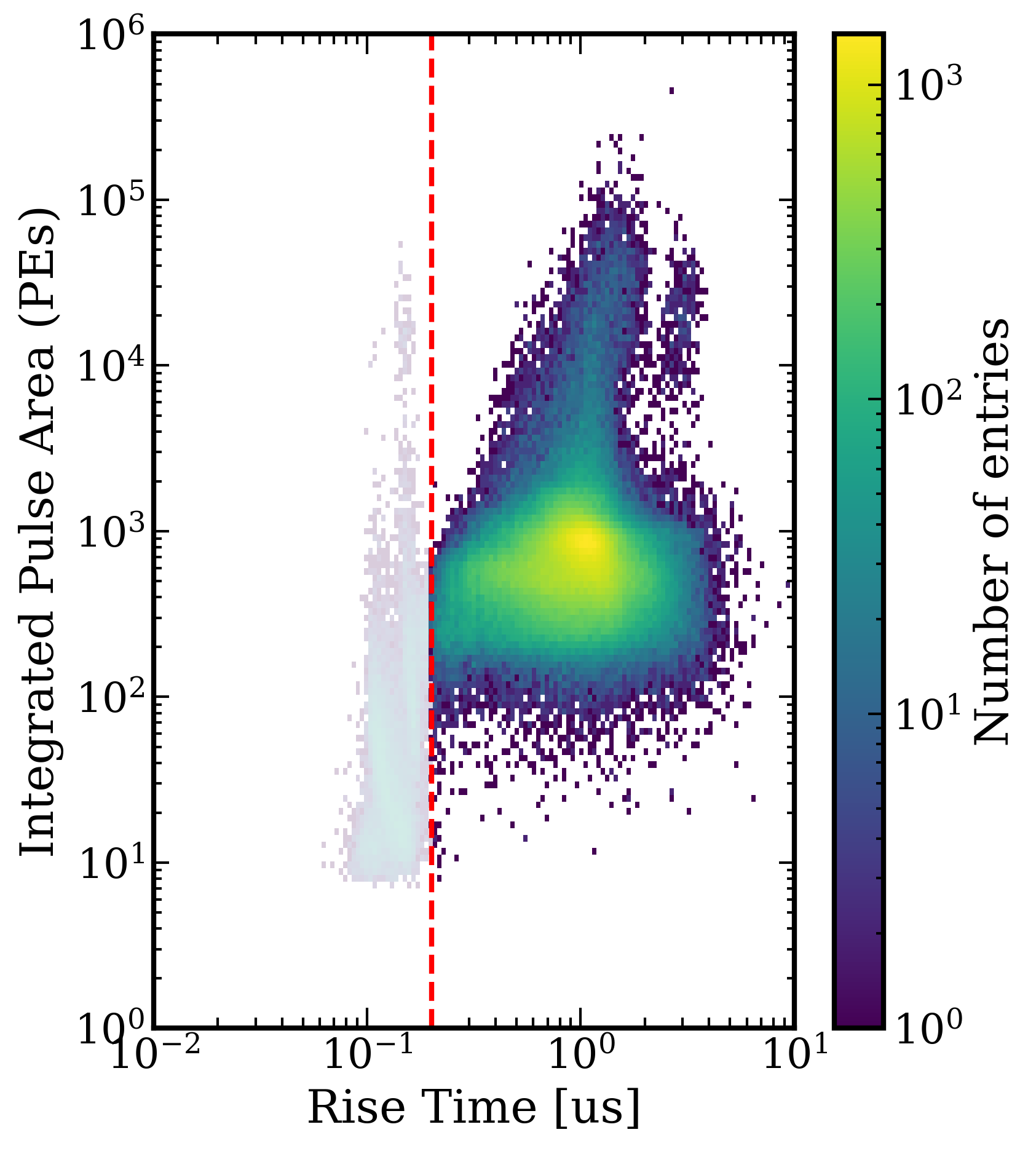}
  \caption{
  The pulse rise time for \iso{241}{Am} events at 4.0\% [Xe] and 10 kV cathode volume as a function of pulse integral as recorded by the sum of all channels. The grayed out region in the left is populated primarily by S1-only or S1-dominated events, due to their faster rise time, and the events with larger rise time are mostly S2 pulses. }
  \label{fig:httCut}
\end{figure}

The largest pulse by area in each remaining event is designated as an S2 candidate and is used to study EL properties in xenon-doped argon. 
At the pulse level, we require a candidate S2 to not have significant contributions from S1s.  
Merged S1-S2 pulses are particularly problematic when the event occurs near the detector perimeter; in these cases the S2 size is significantly degraded, so the event waveform could be dominated by the S1. 
Furthermore, ``S1-only'' events can occur in the reverse-field region under the cathode, where no S2 is generated at all. Both types of events can distort the pulse size and shape in our analysis, but can be identified based on the fact that S1 pulses typically rise to their maximum amplitude at a faster rate than S2s. We quantify this difference with a pulse rise time parameter, defined as the time spanned between when a pulse reaches 50\% and 100\% of its maximum amplitude. An example distribution of pulse rise time and integral in the sum channel for \iso{241}{Am} events is shown in Figure \ref{fig:httCut}. In this figure, S1-dominant events lie in the grayed out region of parameter space, with the S2 events having larger rise times. 
Although S1 and S2 events may overlap significantly in the integral space, they are relatively well separated in pulse rise time.
A 0.2 \us\ cut on the pulse rise time removes nearly all of the S1-dominant population while minimally cutting into the high-energy S2 population of interest, which is verified to hold true across electric fields and xenon concentrations. 

\subsection{Electroluminescence Gain Results}
\label{sec:electroluminescencegainResults}

\begin{figure}[!t]
    \centering
    \includegraphics[width=0.9\columnwidth]{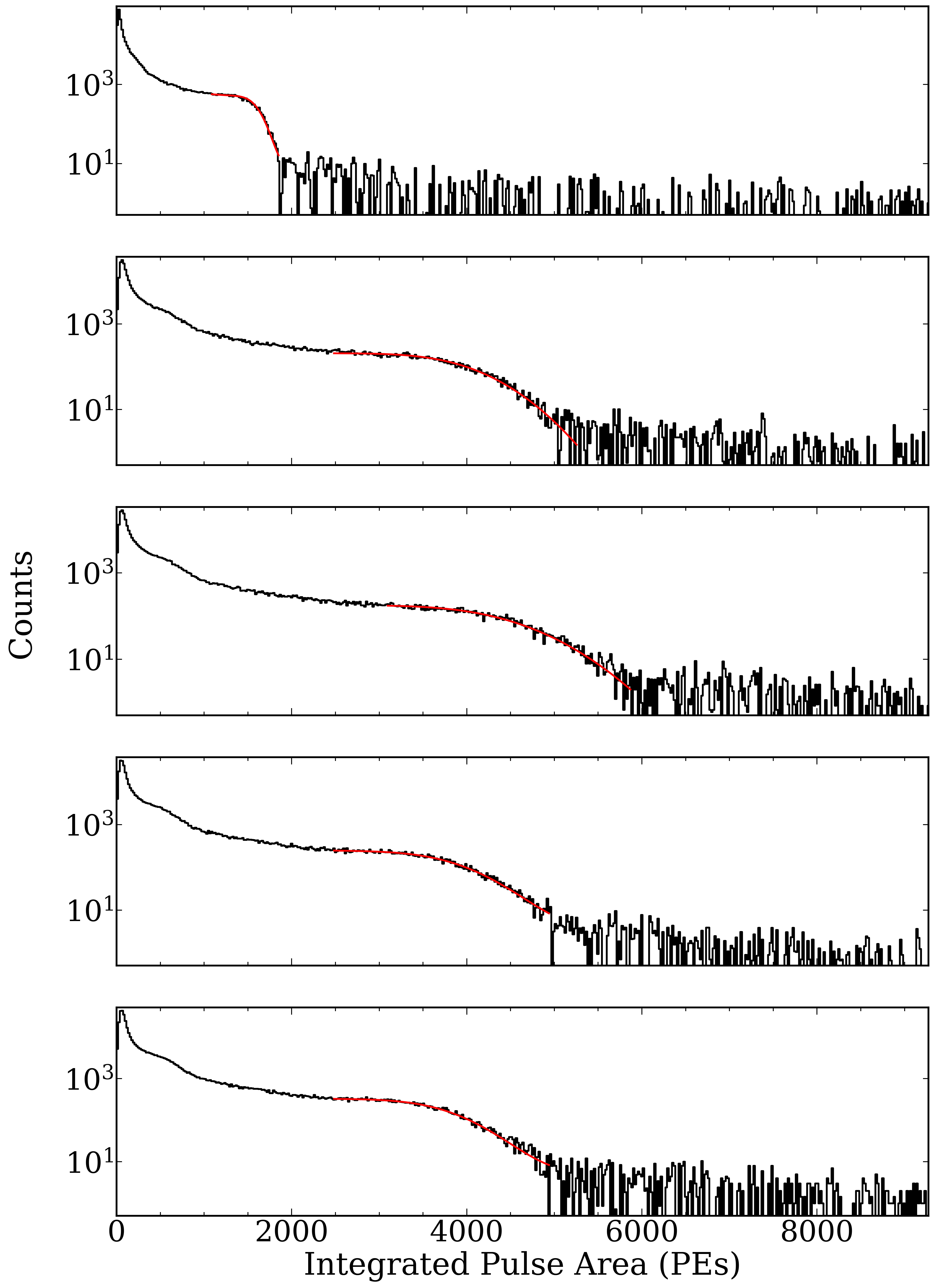}
    \caption{The background-subtracted S2 energy spectra for \iso{241}{Am} calibration events recorded by SiPM Ch 0 at 12~kV for 0--4\% [Xe] (top to bottom). The red curves represent the best fits to extract the endpoint energy (see text).}
    \label{fig:exFits}
\end{figure}

With all aforementioned data quality cuts applied, the remaining events contain S2 pulses produced by \iso{241}{Am} gammas and background interactions by ambient gamma rays and cosmic muons. Contamination from these backgrounds is mitigated by subtracting the livetime-normalized background S2 spectra from those with \iso{241}{Am} taken at the same detector configuration and processed with the same data quality cuts. The resulting \iso{241}{Am} S2 spectrum measured by a top windowless SiPM (Ch 0) for each [Xe] at 12~kV cathode HV is shown in Fig.~\ref{fig:exFits}.

Each \iso{241}{Am} gamma spectrum in Fig \ref{fig:exFits} features a smooth decrease in count rate with increasing S2 size up to an endpoint, beyond which the observed event rate  drops abruptly by a few orders of magnitude. 
This endpoint is attributed to the full-energy \iso{241}{Am} gamma energy deposition right below the SiPM under investigation, where the light collection efficiency (LCE) is the highest. 
For 60 keV gamma rays, approximately 2 in 3 energy depositions in argon are from photoabsorption, with the rest being Compton scatters that have a maximum energy deposition of 11.4 keV. For xenon, the photoabsorption probability is even higher with only 1 in 69 interactions of 60 keV gammas being Compton scattering.
Therefore, the majority of recorded \iso{241}{Am} S2s should contain 60 keV energy deposition across all [Xe] levels in our data, and the low-energy component of the spectrum is explained by the low LCE for events far away from the SiPM channel studied.

We use the endpoint of each spectrum as the metric to characterize S2 sizes in datasets with varying xenon concentrations and electric fields. From pure argon to 1\% [Xe], the endpoint can be seen to increase by a factor of $>$2, suggesting a larger EL yield or an increased SiPM LCE with xenon doping. At higher [Xe] values, the endpoint remains mostly stable. 
To quantify the spectral endpoint, denoted as $E_{0}$, we model the energy spectrum near the endpoint as a Heaviside function $H(E)$ convolved with a Gaussian to capture effects from the finite detector energy resolution. An additional exponential term is used to approximate the small remaining background. 
\begin{align}
    f(E)  = & C_{1}H(E_{0}-E)*\textrm{Gauss}\left(E|\mu=0, \sigma\right) \\ \nonumber
    &+ C_{2}e^{-kE}
    \label{eq:endpoint}
    \end{align}
Each \iso{241}{Am} S2 spectrum is fit to this model using a $\chi^{2}$ minimization. The lower bound of the fit is set where the step function-to-exponential ratio is approximately 1:1 and the upper bound is set where the ratio is approximately 1:3. Representative best-fit functions within the fit range are shown in red in Figure \ref{fig:exFits}. The uncertainty in the endpoint fit from the bound choices is evaluated by varying the bounds and repeating the fits. The lower bound was varied between where the exponential component is approximately twice the step function amplitude and where the exponential contribution is negligible. The upper bound was varied between where the step function contribution is approximately 50\% of the exponential and where the step function component is consistent with 0. 
This study was carried out for each channel for the lowest and highest gain datasets and found to vary the endpoint fit by approximately 1\%. 
Other sources of uncertainty on the endpoint measurement include the SPE calibration and variation of the liquid level. As mentioned earlier, the variation in the SPE calibration over all datasets is less than 2.2\% for all channels. The precision of the level meter allows the gas gap to deviate in length by $\pm$ 7\%, resulting in an electric field that deviates by $\mp$ 1.5\%. An increased gas gap in isolation increases the single electron (SE) gain, while the simultaneous decreased electric field reduces the SE gain per distance~\cite{Oliveira2011_EL}, resulting in a net variation of $\textless$ 5\%.
The combined uncertainty on the measured endpoint values is estimated to be $\textless$ 6\% for all datasets. 

\begin{figure}[!t]
\centering\includegraphics[width=0.95\columnwidth]{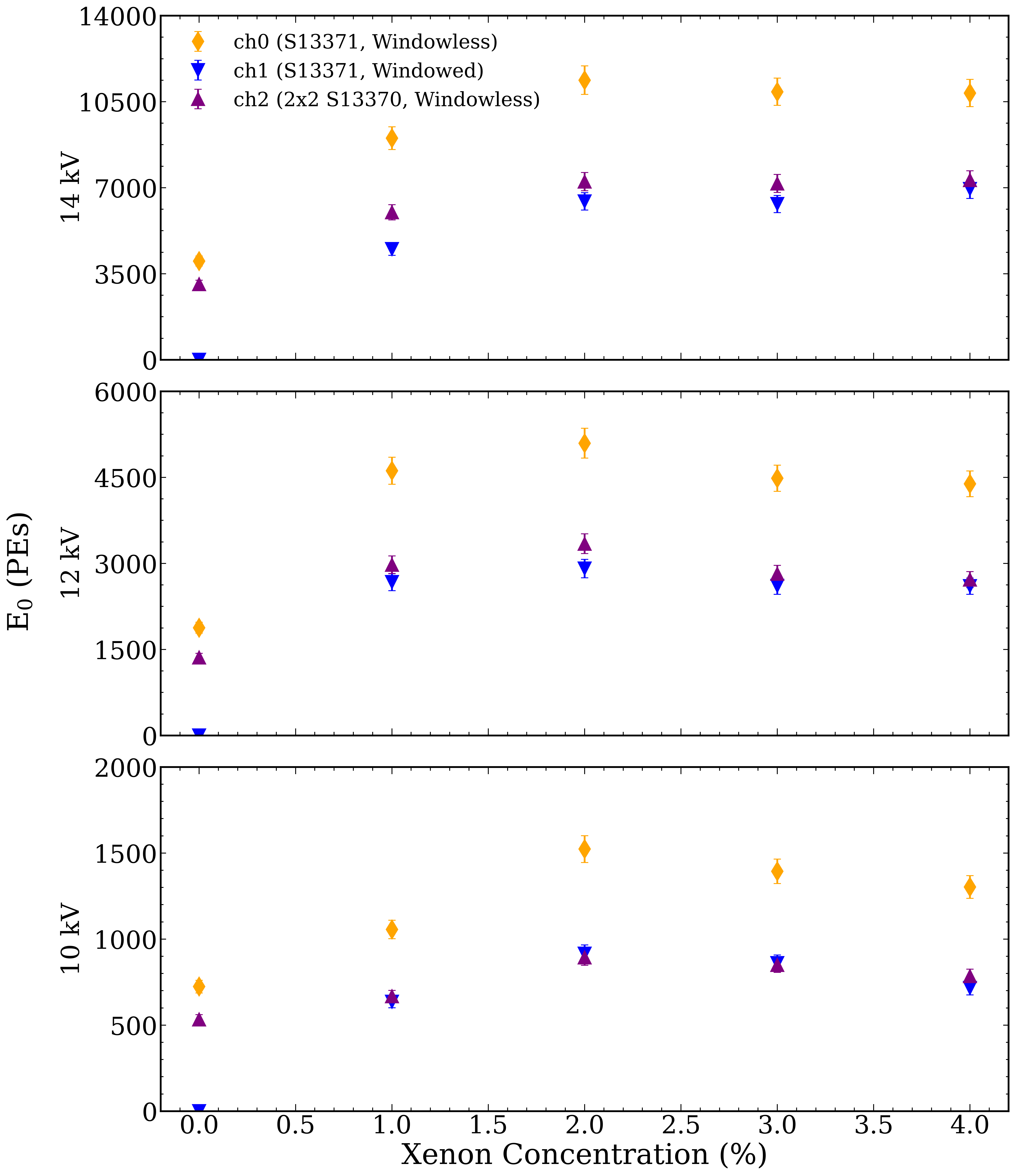}
  \caption{The number of EL photons detected by each top SiPM channel at the endpoint of \iso{241}{Am} gamma energy deposition, as a function of nominal xenon concentration for 3 tested electric fields. The integrated photon area are obtained from fitting the measured S2 energy spectra to a model described in the text.}
  \label{fig:S2GainXeConc}
\end{figure}

Figure \ref{fig:S2GainXeConc} shows the best-fit $E_{0}$ values as a function of the nominal xenon concentration for each SiPM in the top assembly and for each tested cathode voltage. 
The quartz window on SiPM Ch 1 is opaque to 128 nm argon light, causing no photons to be detected in pure argon. At 1\% [Xe], however, the windowed SiPM observes significant light, likely at the Xe$^*_2$ wavelength, with a strength similar to those from the windowless SiPMs. 
The appearance of Xe$^*_2$ EL light also contributes to the increased EL signal size in the windowless SiPMs due to their higher photon detection efficiency (PDE) at longer UV wavelengths. 
For all cathode HV values, $E_{0}$ continues to increase from 1\% to 2\% [Xe], after which $E_{0}$ mostly remains stable. 
This trend is consistent for all SiPMs, suggesting a nearly constant ratio of detected EL light above 160 nm to below. 
The different signal sizes between SiPM channels are attributed to variability in the SiPM PDE value, which may arise from the different pixel sizes of the S13370-6075CN and the S13371-6050CQ-02 SiPMs, the application of different over-voltages, and the effect of the quartz window. 
In addition, SiPM signal crosstalk and afterpulsing may also vary between devices and contribute uncertainties.

\begingroup
\setlength{\tabcolsep}{6pt}
\renewcommand{\arraystretch}{1.2}
\begin{table}[!h]
\centering
\caption{Relative S2 gain factors at the endpoint of \iso{241}{Am} energy deposition for top windowless SiPMs from xenon-doping at 2.125 bar.}
\vspace{6pt}
\label{tab:test}
\begin{tabular}{c|c|c|c|c|c|c}
\multirow{3}{*}{\begin{tabular}{c} HV \\ (kV) \end{tabular}} &
\multirow{3}{*}{\begin{tabular}{c} SiPM \\ CH \end{tabular}} &
\multicolumn{5}{c}{Xenon Concentration} \\
\cline{3-7}
 & & 0\% & 1\% & 2\% & 3\% & 4\% \\
\cline{3-7}
 & & \multicolumn{5}{c}{Relative Gain} \\
\hline
\multirow{2}{*}{14} & 0 &  1 & 2.24 & 2.83 & 2.71 & 2.70 \\
& 2 &  1 & 1.95 & 2.35 & 2.33 & 2.37 \\
\hline
\multirow{2}{*}{12} & 0 &  1 & 2.46 & 2.71 & 2.39 & 2.33 \\
& 2 &  1 & 2.18 & 2.45 & 2.07 & 1.99\\
\hline
\multirow{2}{*}{10} & 0 &  1 & 1.46 & 2.11 & 1.93 & 1.80 \\
& 2 &  1 & 1.25 & 1.68 & 1.59 & 1.47 \\
\end{tabular}
\label{tab:RelativeGainTable}
\end{table}
\endgroup

Table \ref{tab:RelativeGainTable} summarizes the relative S2 gain for both windowless SiPMs from 0\% to 4\% [Xe] at different cathode high voltage values; the windowed SiPM values are not included because no measurement at 0\% [Xe] can be obtained.
At 12 and 14~kV, the highest gain factors are around 2.5. 
Since this gain is accompanied by the appearance of Xe$^*_2$ light detected by the windowed SiPM, we propose that the addition of xenon redirects the energy from Ar$^*_2$ excitations into the emission of longer wavelength photons generated from excitations involving xenon atoms.
At 1\% and 2\% [Xe], we measured gaseous xenon concentrations of $\sim$14 ppm and 34 ppm, respectively, which appear to be effective in facilitating this energy redirection process. Further evidence of this energy transfer phenomenon will be provided in Sec.~\ref{sec:s2_shape}, and more discussions on the increased light yield is provided in Sec.~\ref{sec:disc}.

However, when the xenon concentration in the liquid goes above 2\%, the \iso{241}{Am} S2 sizes appear to decrease slightly. This feature is the most prominent in the 12~kV and 10~kV data. 
The hypothesis of liquid purity degradation over time~\footnote{The typical time interval between doping stages was 1-2 weeks.} was tested by leaving the detector condition unchanged for approximately a week; it was ruled out as a major cause because no significant signal degradation was observed afterwards. 
We explain this change as a possible decrease of electron extraction efficiency from the liquid into the gas as a result of xenon doping. 
Argon has a low electron affinity and it is reported that $\sim$4~kV/cm \cite{eeeGushchin1982HotElectrons} is sufficient to fully extract electrons from liquid into the gas, while xenon requires $>$7~kV/cm \cite{Xu2019_XeEEE}. Therefore, as more xenon is added to liquid argon, the efficiency to extract ionization electrons from the liquid may begin to drop. This effect is partially mitigated in our experiment by the choices of high extraction fields (4.6 - 6.4~kV/cm) but may still be present in the low field datasets. For example, data acquired at 10~kV have the lowest electron extraction field, which also yield the lowest S2 gain at 2\% [Xe] compared to pure argon and experiences the largest drop from 2\% to 4\% [Xe]. %\remark{I think the above paragraph is a mildly conservative argument and belongs here. But we can consider moving it to the discussion.}

\section{Electroluminescence Mechanism}
\label{sec:s2_shape}

In this section, we use the shape of S2 EL pulses in xenon-doped argon to study the EL production process. We first describe the observed changes in the average S2 waveform as xenon is added into liquid argon. Then we build an analytical model to describe the microphysics that transfers energy from argon to xenon and produces EL light.

\subsection{Waveform Shape Observations}
\label{sec:s2_shape_observations}

As xenon is added into liquid argon, the pulse shape of the S2 events is observed to change significantly, but its dependence on xenon concentration appears more complex than that of the S2 size. 
To systematically study the S2 pulse shape, we compute the average S2 waveform at each xenon concentration and cathode voltage for which we took data. By averaging over many events, we suppress the statistical fluctuations in any particular S2 event waveform and obtain the representative timing profile of gaseous argon EL with xenon doping. 
S1 pulses are excluded from the average waveforms using the same cut as described in Sec.~\ref{sec:eventselection}. In addition, we also limit the analysis to events with only one S2 pulse, by rejecting events where the second-largest S2 pulse has an integral over 1\% of the largest one. In this way, the average waveform can be computed over a fixed region of an event window without contamination from multi-scatter events.

For each SiPM channel, we select S2 pulses from full-energy deposition $^{241}$Am events that occur within the line-of-sight of the SiPM channel in question. The origins of these events, the local electric field configuration, and their light collection efficiency are relatively well understood, so the observed changes in the average waveform can be definitively attributed to changes in xenon concentration. 
Specifically, we select $^{241}$Am events near the endpoint of the $^{241}$Am integral distribution (Fig.~\ref{fig:exFits}) to cut against backgrounds from the edge of the detector where the light collection efficiency is poor. For 14~kV on the cathode, this corresponds to approximately over 3000~PE on the top windowless channel.
For each top SiPM, we further select events with at least 45\% of the event's total light seen by the SiPM under study to ensure that the events occur directly under it\footnote{The exact definition of the minimum fraction cut is discussed in Appendix.~\ref{appendix:avg_wf_appendix}. This fraction is, at minimum, 45\%.}. 
The same event selection cuts apply to both windowless and windowed SiPMs in the top assembly, except for the pure argon data in which the windowed SiPM does not detect significant S2 light; in this case events selected for the neighboring top windowless SiPMs are used for the windowed SiPM waveform calculation. 
The bottom SiPMs do not see a substantial portion of the S2 light, so their average waveforms are computed using the same events selected for the top SiPM directly above them.

\begin{figure}[!t]
    \centering
    \includegraphics[width=0.9\linewidth]{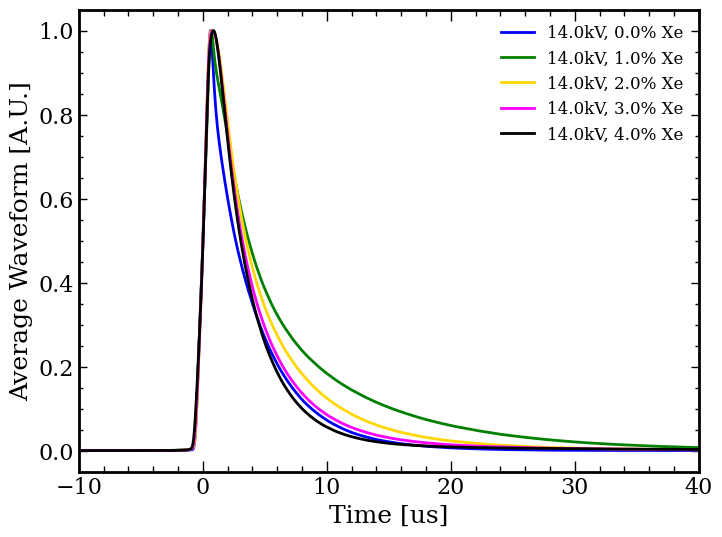}
    \includegraphics[width=0.9\linewidth]{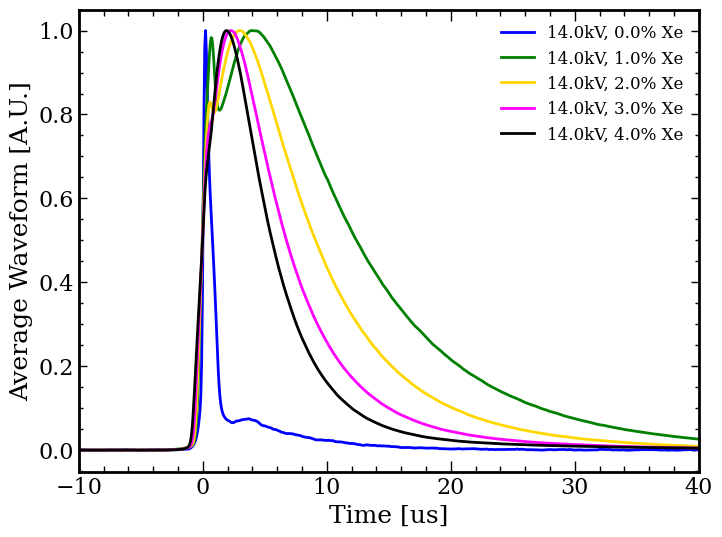}
    \caption{Average S2 waveforms for the windowless (Ch 0, top plot) and windowed (Ch 1, bottom plot) S13371 SiPMs in the top assembly as the xenon concentration in the liquid varies from 0 to 4\% with 14~kV on the cathode. These waveforms are normalized to have a maximum height of 1.
    }
    \label{fig:average_waveforms_14kv_per_channel}
\end{figure}

Figure~\ref{fig:average_waveforms_14kv_per_channel} shows the average waveforms for a top windowless SiPM (Ch 0) and top windowed SiPM (Ch 1) with different [Xe] levels at a cathode voltage value of 14~kV. 
For all SiPM channels, the average S2 waveform at 1\% [Xe] is significantly wider than in pure argon (as measured by the windowless SiPMs). The pulses become narrower again as more xenon is added. At 3--4\% [Xe], the S2 pulses detected by the windowless SiPM
have a width that is nearly the same as the pure argon S2 pulse. 
The changes in the waveform suggest that new energy transfer channels are introduced in xenon-doped gaseous argon during the production of S2 EL light, which will be the focus of the following studies. 

In particular, the windowed SiPM 
experiences drastic changes in the S2 pulse shape as [Xe] rises from 0 to 4\%. 
In pure argon, the S2 pulses in the windowed SiPM are substantially narrower than those of the windowless SiPM. Since argon EL light is released in several microseconds, the lack of a characteristic argon EL tail confirms that the windowed SiPMs are insensitive to photons from Ar$_2^*$ light, with a wavelength around 128~nm. However, the presence of the narrow prompt peak suggests that photons with wavelengths greater than the quartz cutoff wavelength of 160~nm are emitted as the electrons traverse the gas under a strong electrical field. Such an emission of light is consistent with neutral bremsstrahlung as reported in \cite{Buzulutskov:2018vgg}, which consists of rapidly emitted broad-spectrum light.

\begin{figure}[!htb]
    \centering
    \includegraphics[width=0.9\linewidth]{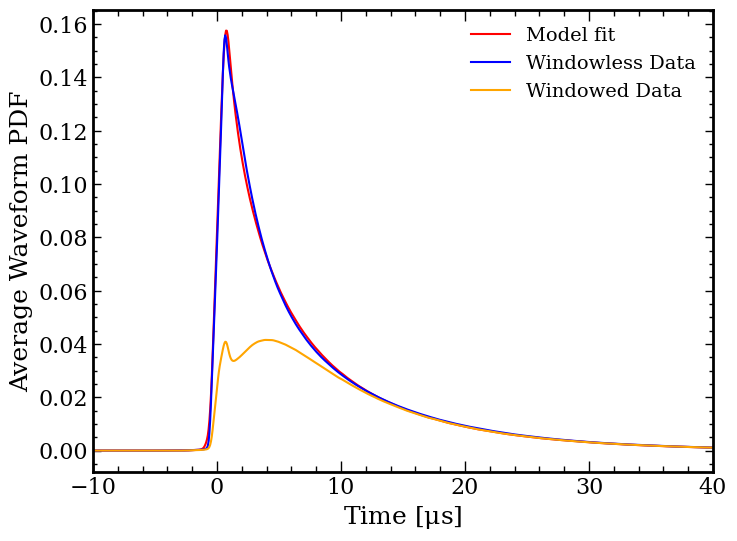}
    \includegraphics[width=0.9\linewidth]{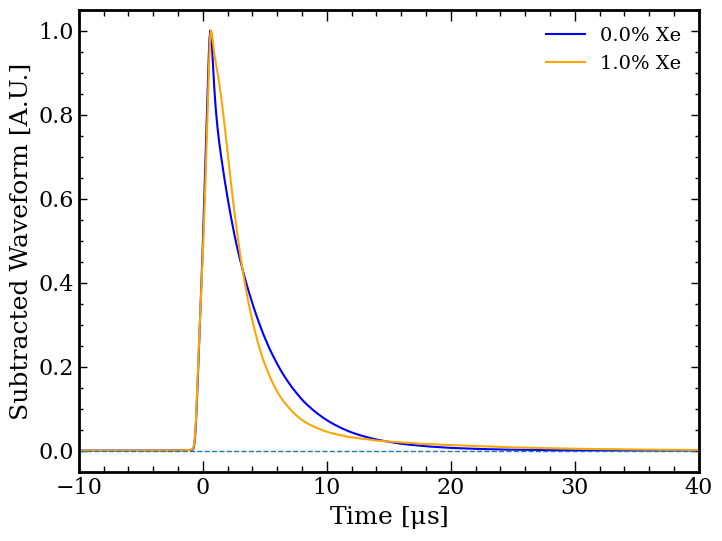}
    \caption{\textbf{Top}: Average waveforms for the top windowed (Ch 1) and windowless (Ch 0) SiPMs at 1\% [Xe] and 14~kV on the cathode. A fit to the Ch 0 waveform including both the Ch 1 waveform template and additional Ar$^*_2$ components (elaborated in in Sec.~\ref{sec:modeling_s2}) is also shown. \textbf{Bottom}: the resulting waveform from subtracting the scaled Ch 1 waveform from the Ch 0 waveform. This subtracted waveform estimates the time profile of the Ar$_2^*$ light by itself.}
    \label{fig:ch0_sipm_avgwf_fit}
\end{figure}

More interestingly, at 1\% [Xe], the Ch 1 waveform contains a second hump after the prompt peak. This second hump becomes sharper and moves to earlier times as more xenon is added. 
Given that the top windowless SiPM is similarly sensitive to $>$160nm photons, we hypothesize that the same double-hump structure is also present in the Ch 0 waveform, but is obscured by the additional detection of 128~nm Ar$_2^*$ EL light. In Fig.~\ref{fig:ch0_sipm_avgwf_fit} (top), we show that the tails of the top windowed and windowless SiPMs' average waveforms at 1\% [Xe] can be matched, with the windowed SiPM waveform exhibiting a deficit at small times. 
The difference between these two SiPMs' average waveforms is attributed to light emitted from Ar$_2^*$, and is only visible to the windowless SiPM~\footnote{As discussed later, some Xe or ArXe excimers may also emit $<$160~nm photons, but such processes will be slow and their contributions are expected to be similar to that detected by the windowed SiPM.}.
Fig.~\ref{fig:ch0_sipm_avgwf_fit} (bottom) shows the subtracted Ar$_2^*$ waveforms obtained by taking the difference between the windowed and windowless SiPMs' waveforms with proper scaling (see Appendix~\ref{appendix:avg_wf_appendix}). 
The observed tail of the Ar$_2^*$ component of the S2 waveform at 1\% [Xe] is shorter than that for pure argon, indicating that energy is diverted away from Ar$_2^*$ to Xe: the energy transfer to Xe competes with Ar$_2^*$  de-excitation, and thus Ar$_2^*$ light can only be observed if the decay occurs \textit{before} this energy transfer.

The second hump in the top windowed SiPM waveform can then be explained as light produced from the de-excitation of Xe$_2^*$. The formation of Xe$_2^*$ requires Ar$_2^*$ to first transfer its energy to a Xe atom, after which the excited Xe needs to find another Xe atom to form Xe$_2^*$ before promptly emitting 175~nm light. The rate of both processes should be proportional to the gaseous xenon concentration -- which is only several tens of ppm -- and are thus slow.
The rising edge of the second hump corresponds to the buildup of excited Xe, while the slow decay can be attributed to the formation and subsequent de-excitation of Xe$_2^*$ into a 175~nm photon. As the xenon concentration increases, both processes take less time on average, thus narrowing the S2 pulse. Such a hump has been observed before in liquid-phase xenon doped argon experiments as well \cite{Galbiati:2020eup, Wahl:2014vma, X-ArT:2024npd, DUNE:2024dge}, and is quantitatively modeled in the following section.

\begin{figure}[!htbp]
    \centering
    \includegraphics[width=0.9\columnwidth]{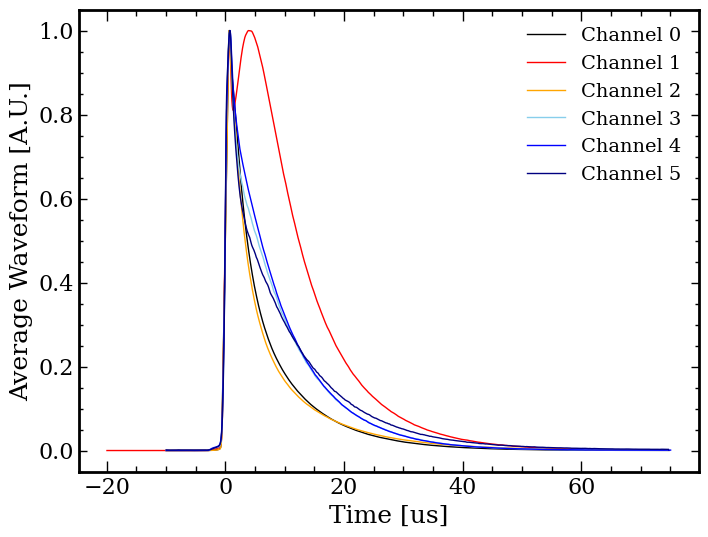}
    \caption{The average waveforms across all channels for 1\% xenon doping in liquid argon. Chs 3, 4, and 5 are on the bottom of the TPC and their average waveforms are computed using events below their corresponding top assembly SiPMs. The presence of wavelength shifting from Ar$_2^*$ light to Xe$_2^*$ light in the liquid gets rid of the double-hump structure on the bottom windowed SiPM.
    }
    \label{fig:all_wfs_1p_xe}
\end{figure}

Unlike the top windowed SiPM in the gas, the average waveform for the windowed SiPM in the liquid (Ch 4), as shown in Fig.~\ref{fig:all_wfs_1p_xe}, does not have a double-hump structure. Furthermore, its waveform shape is largely similar to those of its neighboring bottom windowless SiPMs. The $^1P_1$ and $^3P_2$ states of atomic Xe have energies around 129.56~nm and 146.96~nm wavelengths~\cite{GNowak_1985, 10.1063/1.2136879}, which may make the xenon-rich liquid highly absorbent to 128~nm and 147~nm light, according to the NIST Atomic Spectra Database \cite{NIST_ASD}.
Thus, we propose that these short wavelength photons are efficiently absorbed and wavelength-shifted to 175~nm when they reach the xenon-rich liquid.
%In addition, this wavelength-shifting phenomenon can also be inferred by examining the shape of the average S2 waveform on the top windowed SiPM for events that do \textit{not} occur directly beneath the SiPM in question, as discussed in Appendix~\ref{appendix:avg_wf_appendix}. 

\subsection{Modeling of Xenon-doped Argon Electroluminescence}
\label{sec:modeling_s2}
Based on the aforementioned observations, we construct an approximate model to describe the energy transfer from Ar to Xe and to explain the leading-order changes in the S2 pulse shape as a result of xenon doping.
Discussions of higher-order effects can be found in Appendix~\ref{appendix:avg_wf_appendix}. 
An S2 pulse is produced by ionization electrons that drift and diffuse in the liquid, are extracted into the gas, and then create excited atoms. These excited atoms then form excited molecules, called excimers, which de-excite and release VUV photons. In pure argon or pure xenon, these excimers are widely known as Ar$_2^*$ and Xe$_2^*$, respectively. For an event at a given detector depth $z$, the processes above can be modeled as the following set of convolutions, \begin{equation}
\begin{split}
    P_{S2}(t) & = \text{Gauss}\left(t|\mu=\frac{z}{v_d}, \sigma =\sqrt{\frac{2D_Lz}{v_d^3}}\right)\\ 
    &*f_{SE}(t)*P_{light}(t)*f_{SPE}(t)
\end{split}
\label{eq:full_s2_pulse}
\end{equation} 
The Gaussian term accounts for the diffusion of the electron cloud as it drifts through the liquid, where $z$ is the event depth as measured below the liquid surface, and $D_L$ is the longitudinal diffusion constant. $f_{SE}(t)$ is the distribution of times at which an excitation is created during a single electron's drift from the liquid surface to the TPC anode, and $f_{SPE}(t)$ is the average shape for a single photoelectron pulse.  Finally, $P_{light}(t)$ is the distribution of time for which photons are released in the de-excitation chain, as described below.

As the xenon concentration in the gas in our experiment is only tens of ppm, collisions of accelerated electrons will primarily excite Ar atoms, producing Ar$^*$. Ar$^*$ will then find a ground state Ar to form a singlet or triplet dimer, denoted Ar$_2^{*1}\Sigma$ and Ar$_2^{*3}\Sigma$. The time-scale of this dimer formation in gaseous argon is estimated to be less than a nanosecond based on a density scaling of the picosecond-scale dimer formation time in liquid argon \cite{LORENTS197619}. 
In pure argon gas, the singlet de-excites in $\sim$4.2~ns, while the triplet de-excites in $\sim$3.3~$\mu$s \cite{PhysRevLett.33.1365}. 
The addition of xenon to argon gas gives rise to new pathways of energy transfer and causes the deexcitation pulse shape to deviate from that of pure argon, which have also been seen in many other experiments examining the scintillation of gaseous xenon-argon mixtures via spectroscopy \cite{Takahashi1975_Xe, 10.1063/1.1680473, 10.1063/1.434079}.

\begin{figure}[!htbp]
    \centering
    \includegraphics[width=0.9\columnwidth]{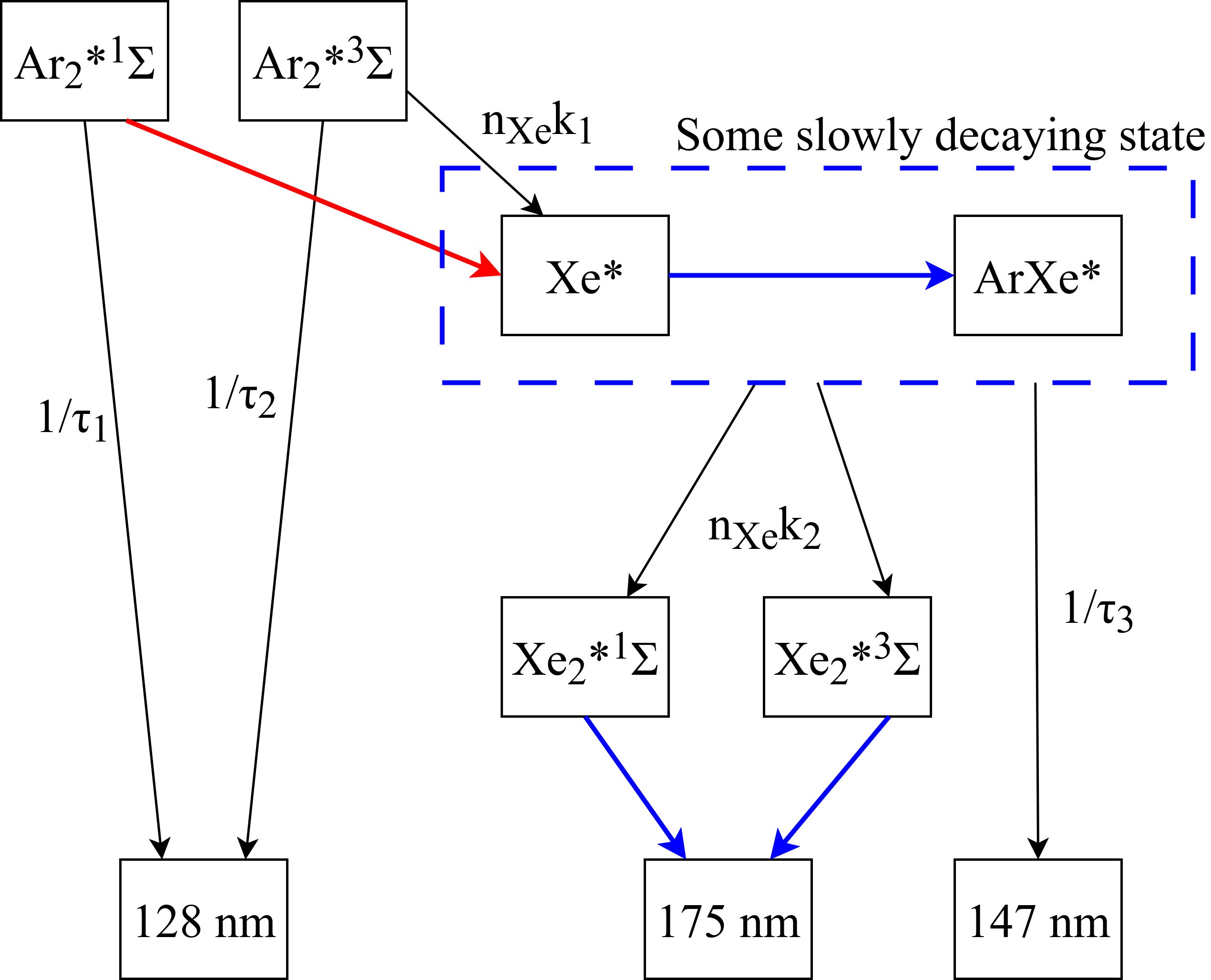}
    \caption{A hypothesis for the scintillation mechanisms in a gaseous mixture of xenon in argon. Ar$_2^*$ singlets forming Xe$^*$ are unlikely to occur since the Ar$_2^*$ singlet would have probably decayed beforehand (shown in red). The blue arrows are assumed to be too fast ($<$1~$\mu$s) and their decay times are therefore neglected. The dashed blue box indicates that it is unclear whether or not ArXe$^*$ is formed.}
    \label{fig:arxe_scintillation_gas}
\end{figure}

We propose leading-order pathways for de-excitation in gaseous xenon doped argon as illustrated in Fig.~\ref{fig:arxe_scintillation_gas}. 
An Ar$_2^*$ dimer can either de-excite by emitting a 128~nm photon, or collide and transfer its energy to a ground state Xe atom to form an excited Xe$^*$. A collision between Ar$_2^*$ and Xe is only likely to happen if Ar$_2^*$ is in the triplet state, as the singlet state decays far too quickly. Xe$^*$ may \cite{Galbiati:2020eup, KUBOTA199371, Neumeier_2015, GNowak_1985} or may not \cite{TAKAHASHI1983591, 10.1063/1.434079} go on to form ArXe$^*$, as the literature is unclear. If Xe$^*$ does form ArXe$^*$, it will likely do so quickly since Ar is abundant in the gas. In any case, such a state, whether Xe$^*$ or ArXe$^*$, has been observed to decay by emitting a $\sim$147~nm photon according to previous wavelength spectrum measurements \cite{TAKAHASHI1983591, 10.1063/1.434079}. However, Xe$^*$ or ArXe$^*$ is reported to be long-lived \cite{10.1063/1.434079}, which then allows the formation of Xe$_2^*$ and the emission of 175~nm light. 
Otherwise, the energy will be primarily dissipated as 147~nm photons, and the top windowed SiPM would not have observed significant Xe$_2^*$ de-excitation S2 light.

Quantitatively, the formation of Xe$_2^*$ involves Ar$_2^{*3}\Sigma$ forming Xe$^*$ with a rate of $k_1 n_{Xe}$, followed by the long-lived intermediate state forming Xe$_2^*$ with a rate of $k_2n_{Xe}$. Here, we have assumed that the energy transfer rates are proportional to the number density of Xe in the gas phase, $n_{Xe}$, which will be validated with our experimental observations later in this section. Since this is a two-step exponential process, the time profile can be described as the convolution of two exponentials with time constants $\tau_A$ and $\tau_B$, which is generically of the form \begin{equation}
    \frac{1}{\tau_A-\tau_B}(e^{-t/\tau_A}-e^{-t/\tau_B})\label{eq:twoexp}
\end{equation} and gives rise to the observed rounded second hump structure.

With these assumptions, the shapes of the S2 waveforms can be calculated as follows. Excitation collisions from electrons create some number of Ar$_2^{*1}\Sigma$ singlets, $N_{Ar}p_{S}$, where $N_{Ar}$ represents the total number of Ar excimers, and $p_S$ is the proportion of singlet excimers, so the number of triplets becomes $N_{Ar}(1-p_S)$. From here, the time profile of Ar$_2^*$ de-excitation is then \begin{align}
    P_{Ar}(t) & = N_{Ar} \left(p_S\frac{1}{\tau_1} e^{-t/\tau_1} \right.\label{eq:ar_timing} \\ & \nonumber \left.+ (1-p_S)\frac{1}{\tau_2} e^{-(\frac{1}{\tau_2} + k_1 n_{Xe})t}\right)
\end{align} where the rate constant $k_1$ is defined~\footnote{For a given relative velocity between two atoms ($\vec{v}$) and interaction cross section ($\sigma(\vec{v})$), $k = n_{Xe} \sigma(\vec{v})|\vec{v}|$.} in Fig.~\ref{fig:arxe_scintillation_gas}. $\tau_1$ and $\tau_2$ are the time constants for the singlet and triplet decay in pure argon. Note that the rate of triplet de-excitation is shortened by the rate at which Ar$_2^{*3}\Sigma$ forms Xe$^*$. This describes the narrowing of $P_{Ar}(t)$ in Fig.~\ref{fig:ch0_sipm_avgwf_fit} (bottom). 

After Xe$^*$ or ArXe$^*$ is formed, it can either decay into a 147~nm photon, or form Xe$_2^*$, with a corresponding rate of $1/\tau_3$ and $k_2n_{Xe}$, respectively. 
These two competing processes in the second step cause the emission of 147~nm light and the formation of Xe$_2^*$ to share the same time profile with a combined decay rate of $k_2n_{Xe}+1/\tau_3$. 
In our experiment, this time profile can be measured with the second hump in the top windowed SiPM waveform, which follows the general prescription of Eq.~\ref{eq:twoexp}. 
This time profile for Xe$_2^*$ is modeled as
\begin{align}
    P_{Xe}(t) &= \frac{N_{Ar}(1-p_S)k_1 n_{Xe}k_2 n_{Xe}}{\frac{1}{\tau_2} + k_1 n_{Xe} - \frac{1}{\tau_3} - k_2 n_{Xe}} \label{eq:xe_timing} \\& \nonumber \times\left(e^{-t(\frac{1}{\tau_3} + k_2 n_{Xe})} - e^{-(\frac{1}{\tau_2} + k_1 n_{Xe})t}\right).
\end{align}
For the 147~nm light emission, the formula only differs in the coefficient: 
\begin{align}
    P_{147}(t) & = \frac{N_{Ar}(1-p_S)k_1 n_{Xe}\frac{1}{\tau_3}}{\frac{1}{\tau_2} + k_1 n_{Xe} - \frac{1}{\tau_3} - k_2 n_{Xe}} \label{eq:147nm_timing} \\& \nonumber \times \left(e^{-t(\frac{1}{\tau_3} + k_2 n_{Xe})} - e^{-(\frac{1}{\tau_2} + k_1 n_{Xe})t}\right).
\end{align}
The relatively fast de-excitation of Xe$_2^*$ ($\sim$10~ns) is neglected in our model.
It is readily apparent that Eq.~\ref{eq:xe_timing} predicts an initial broadening and then narrowing of the slow hump in the top windowed SiPM's S2 waveform as xenon is added to Ar. 

\begin{figure}[!hb]
    \centering
    \includegraphics[width=0.9\columnwidth]{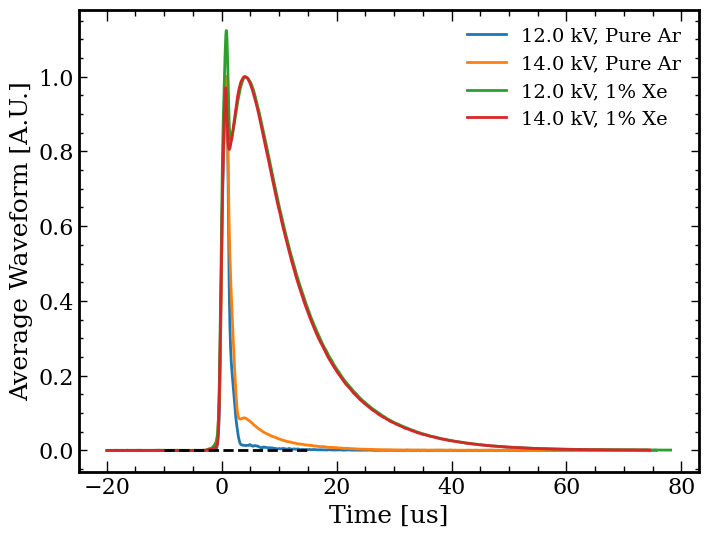}
    \caption{A comparison of the top windowed SiPM's average waveforms for 0\% and 1\% xenon in liquid argon, and for 12~kV and 14~kV on the cathode. The fast hump is shown to be of the same shape across all configurations, and is more prominent for lower voltages. The 0\% (1\%) [Xe] waveforms are normalized to the height of the first (second) hump.}
    \label{fig:neutral_brem_hypothesis}
\end{figure}

A full waveform model also requires a description of the first hump observed in the top windowed SiPM average waveform, which we attribute to neutral bremsstrahlung. Neutral bremsstrahlung occurs when a drifting electron scatters elastically off a bound electron in an atom, and produces broad-spectrum light. This process has been observed before in a dual phase argon TPC \cite{Buzulutskov:2018vgg}, and can happen at reduced electric fields \textit{lower} than the onset of excimer EL. Past the excimer EL threshold field, the predicted photon yield from neutral bremsstrahlung quickly plateaus~\cite{Buzulutskov:2018vgg}, while the excimer EL photon yield continues to grow. As such, neutral bremsstrahlung is expected to be more prominent at lower reduced electric fields, which is consistent with our observations.  Fig. \ref{fig:neutral_brem_hypothesis} shows the average waveform in the top windowed SiPM computed with data taken at 12~kV and 14~kV on the cathode; when both waveforms are normalized using the slow hump, the one taken at lower reduced elected field (12~kV) has a more prominent prompt component.

%=======================
% MAIN POINTS:
% - Model can fit the shape quite well (see figures in appendix)
% - Model also has a fairly linear fit to the rates of Xe* and Xe_2^* production as a function of Xe concentration
% - Model only explains first order phenomena, many unexplored phenomena are discussed in the appendix

\begin{figure}[!t]
    \centering
    \includegraphics[width=0.9\linewidth]{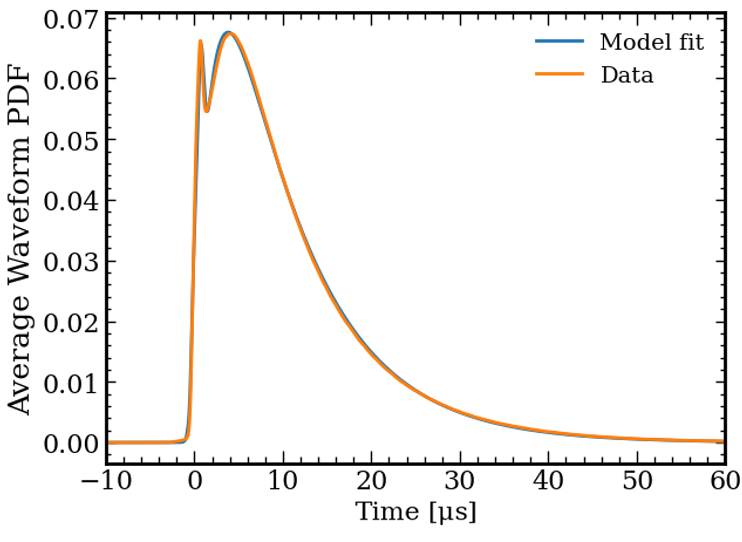}
    \includegraphics[width=0.9\linewidth]{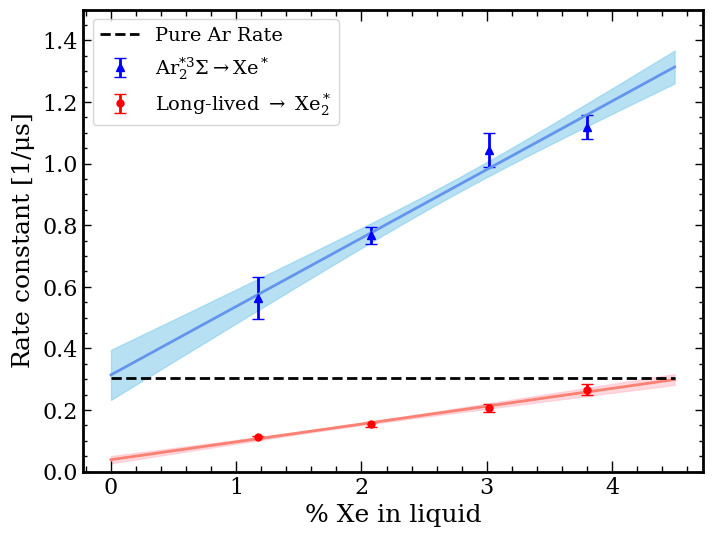}
    \caption{\textbf{Top}: an example average S2 waveform fit for Ch 1 at 1\% [Xe] in the liquid. \textbf{Bottom}: the rates of Ar$_2^{*3}\Sigma$ to Xe$^*$, and the metastable state to Xe$_2^*$ as a function of [Xe] in liquid as obtained from fitting the EL model to the top windowed SiPM. We see a clear linearity of these rates with [Xe]. The black dashed line corresponds to the reciprocal of the 3.3~$\mu$s Ar$_2^{*3}\Sigma$ lifetime from \cite{PhysRevLett.33.1365}.}
    \label{fig:k1_k2_linearity}
\end{figure}

By treating the average waveforms as a linear combination of the different excimer de-excitation components as well as the prompt light released by neutral bremsstrahlung, we fit our EL model to data. 
Examples of fitted average waveforms for a top windowed (Ch 1) and windowless (Ch 0) SiPM 1\% [Xe] are shown in the top subfigures of Fig.~\ref{fig:k1_k2_linearity} and Fig.~\ref{fig:ch0_sipm_avgwf_fit}, respectively.
The full fit results over all measured [Xe] values can be found in Fig.~\ref{fig:windowed_sipm_avgwf_fits} and Fig.~\ref{fig:windowless_sipm_avgwf_fits} in Appendix~\ref{appendix:avg_wf_appendix}. %, where further details regarding the fit are also discussed. 
This data-model agreement quantitatively explains our major observations, such as the double-hump structure of the top windowed SiPM waveform and the widening and subsequent narrowing of the S2 pulse with increasing [Xe]. 

Furthermore, the fits also provide an estimate of the rate of Xe$^*$ formation from Ar$_2^{*3}\Sigma$, as well as the rate of Xe$_2^*$ formation from either Xe$^*$ or ArXe$^*$, both of which are shown as a function of xenon concentration in Fig.~\ref{fig:k1_k2_linearity} (bottom). 
The rate coefficients are observed to increase linearly with [Xe], as predicted by our model. 
Notably, the intercepts of these lines at 0\% [Xe] are not 0, corresponding to the intrinsic decay times of Ar$_2^{*3}\Sigma$ and Xe$^*$ (or ArXe$^*$) in the absence of interactions with Xe. 
Further details on the data-model comparison can be found in Appendix~\ref{appendix:avg_wf_appendix}, where higher-order effects, such as the relative quantities of each wavelength component and additional de-excitation pathways are also discussed. 
%However, there are still some unexplained phenomena regarding gaseous xenon-doped argon EL that are beyond the scope of this model. These include the relative quantities at which 128~nm, 147~nm, and 175~nm light are produced, as well as the potential presence of other de-excitation pathways not included in Fig.~\ref{fig:arxe_scintillation_gas} that may give rise to a long, slowly decaying tail in the S2 pulse. These remaining questions, and our current hypotheses for the origins of these phenomena are discussed 
\iffalse
\begin{figure}[!htbp]
    \centering
    \includegraphics[width=0.9\linewidth]{just_ch1_wf_fit.png}
    \caption{An example average S2 waveform fit for Ch 1 at 1\% [Xe] in the liquid.}
    \label{fig:1pxe_only_windowed_wf}
\end{figure}
\fi

\section{Discussion}
\label{sec:disc}

The most important observation in this work is that the presence of xenon in gaseous argon is efficient in converting Ar$_2^*$ EL into the Xe$_2^*$ wavelength. 
This interpretation is supported by the large increase of S2 size recorded by the top windowless SiPMs, which is accompanied by the detection of significant Xe$^*_2$ light by the top windowed SiPM (Sec.~\ref{sec:electroluminescencegainResults}) and the energy transfer from Ar$^*_2$ to Xe indicated by changes in S2 waveform (Sec.~\ref{sec:s2_shape}). 
The energy transfer from Ar excitons to Xe in the liquid has been extensively studied, and it was reported that $\mathcal{O}$(10)ppm of xenon doping in the liquid can shift a significant fraction of liquid argon scintillation light into the Xe$_2^*$ wavelength. 
Our work demonstrates that the efficiency of energy transfer in the gas during EL is comparable to that in the liquid. 
At 1\% [Xe], we measure $\sim$14 ppm of xenon in the gas phase, and the observed EL already contains $\sim$50\% 175~nm Xe$_2^*$ emission. 

In addition, we report that liquid argon with $\ge$1\% xenon doping is a potent wavelength shifter for VUV photons. This is evidenced by the nearly identical photon signals recorded by the SiPMs with and without quartz windows operated in the liquid mixture (Fig.~\ref{fig:all_wfs_1p_xe}). Quartz does not transmit VUV photons with wavelengths below 160~nm, so the windowed SiPM should only be sensitive to 175~nm photons from Xe$_2^*$ emission. 
However, both the average waveform and the integrated S2 spectra measured by the windowed-SiPM in the liquid are consistent with those by nearby windowless SiPMs. 
This degeneracy indicates that photons arriving at the bottom SiPM assembly are essentially all above the quartz cutoff wavelength and are from Xe$_2^*$ deexcitations. 
To the contrary, data recorded by SiPMs in the gas show distinct waveform shapes and pulse integrals. 
Therefore, we conclude that the EL light emitted in the gas consists of photons both above and below the quartz cutoff wavelength, but nearly all of $<$160~nm photons are shifted to longer wavelengths after they pass the 1~cm of liquid argon with 1\% xenon doping. 

As the liquid [Xe] is increased from 1\% to 4\%, the measured xenon concentration in the gas also increases from 14 ppm to 61 ppm. 
Consequently, the energy transfer and light emission processes for gas EL undergo significant changes, as reflected by the fast evolving pulse shape observed by both the windowed and windowless SiPMs in the gas (Fig.~\ref{fig:average_waveforms_14kv_per_channel}). 
However, the significant pulse shape changes, or the enhanced energy transfers, are not accompanied by a proportionate change in the total number of photons collected. 
This observation may be partially attributed to a slight decrease in the efficiency of extracting ionization electrons from the liquid into the gas (Sec.~\ref{sec:electroluminescencegainResults}). Although each extracted electron may produce a larger EL signal, the increase is negated by the smaller number of electrons extracted. 
In addition, the apparently stagnant ratio of 175~nm light to that of total light could be explained by the small contribution from liquid wavelength-shifted 128~nm and 147~nm light, which anti-correlates with the production of 175~nm light and can offset its effect.

%\remark{Inserting paragraph about boost to light yield here. Also moved around some sentences at the end of this and the next paragraph to flow better.} As importantly, we observe a significant relative S2 gain improvement across all SiPM channels. 
We note that the observed S2 gain improvement cannot be fully explained by the energy transfer from argon to xenon in the EL process with xenon doping. 
Even if every 128~nm argon EL photon is replaced with a 175~nm photon, the observed photon signal in the windowless channels should only increase to $\sim$1.7 times the pure argon result. 
This is estimated from the 67\% higher photon detection efficiency of VUV4 SiPMs at 175~nm than at 128~nm, as reported by Hamamatsu~\cite{HamamatsuVUVGraph}.
However, a substantially higher gain is seen in most windowless channels datasets for xenon concentrations greater than 1\%.
We attribute the additional S2 gain to other processes related to xenon doping. For example, the presence of xenon in the gas phase can cause some xenon to be directly excited. This can increase the EL yield itself due to the lower first excited energy of xenon (8.3~eV) relative to argon (11.5~eV)~\cite{RevModPhys.82.2053}, increasing the size of EL signals for each extracted ionization electron.  
Additionally, the lower ionization energy of xenon compared to that of argon means that each ionization on xenon requires less energy than for argon so more ionization electrons may be produced for the same energy deposited in the liquid when percent-level xenon is present. 

The liquid argon ionization yield enhancement with xenon doping is expected to be even higher for low-energy nuclear recoils than for electron recoils. 
Nuclear recoils mainly dissipate energy through scattering with the nuclei of neighboring atoms, and this cascade process would produce a population of nuclear recoils heavily tilted toward low energies. Therefore, if xenon doping indeed lowers the effective ionization energy of liquid argon, this effect can cause a disproportionate increase of ionization yields for argon nuclear recoils. Combined with the high efficiency of collecting Xe$_2^*$ EL light, the performance of a dual-phase xenon-doped argon detector can surpass that of either argon or xenon for low-energy nuclear recoil detection.

The observed number of photons in this experiment is subject to various uncertainties in addition to statistical fluctuations, such as small drifts of detector liquid level and device-to-device variability in the SiPM PDE. 
First, to mitigate the risk of xenon segregation in a conventional purification scheme that boils liquid from a spill-over ``weir'' for gas purification, we directly circulate boil-off gas from the detector volume. The lack of a weir allows the liquid level to drift, for example, due to a change in liquid partition between the detector volume and the condensation volume. Small oscillations in the liquid level over a time scale of days were observed during otherwise stable operation when no changes in operation parameters were enacted. We address this issue by checking the liquid level prior to every data acquisition but the measurement mechanism is discrete and allows for $<$1.2~mm liquid surface variability. 
This liquid level effect is found to contribute to $\sim$5\% uncertainty to our S2 pulse size measurements. 
Second, we choose 3 SiPM modules of different characteristics for both the top and bottom SiPM assemblies to maximize information that can be extracted from experimental observations. 
However, this also means that these 3 SiPM modules can have different PDE values for photons of the same wavelength and different relative PDE ratios for photons of different wavelengths. 
These SiPMs can also have different afterpulsing and cross-talk probabilities. As a result, we refrain from drawing quantitative conclusions from the comparison of signal strengths between different SiPMs. 
These uncertainties are not significant enough to challenge the interpretation of observed data. 

The main analyses presented in this work are based on measurements at a detector pressure of 2.125 bar. 
The high operating pressure was chosen to improve xenon solubility in liquid argon and also xenon vapor pressure, both of which increase with system temperature. 
However, the experimental observations suggest that 1--2\% of xenon doping in liquid argon may provide the strongest enhancement of S2 gain, which opens up more parameter space for detector operations. 
In fact a lower detector pressure can provide some appealing benefits for S2 gain.
For example, even with the same electric field, a lower gas pressure leads to a higher reduced electric field and a higher S2 gain. 
As part of this work, we also studied the S2 signals in pure argon and in 1\% xenon-doped argon at a lower pressure of 1.75 bar. 
In both pure argon and 1\% [Xe]  the measured S2 gain values are higher at 1.75 bar than those at 2.125 bar for the same electric field. 
However, at 1.75 bar, the gaseous xenon concentration is lower than that at 2.125 bar, leading to a smaller relative S2 gain of $\sim$2 from 1\% xenon doping. 

In addition, at 1.75 bar operating pressure, the high voltage system of the TPC exhibited significant instabilities at 1\% xenon doping, signified by sudden increases in the photon signal rate and difficulties in maintaining high voltage values. 
As a result, we were unable to operate the detector stably at 14~kV on the cathode at 1\% xenon doping while the system could maintain 20~kV in pure argon or 14~kV in 4\% xenon-doped argon (for short durations) at 2.125 bar. As a result xenon doping of $>$1\% levels was not attempted at 1.75 bar. 
The exact cause of the high voltage instability is unknown, but appears to be related to the presence of large quantities of xenon in the liquid. This was confirmed by a test in which we raised the liquid level to above the TPC anode (grounded) so no gas electroluminesence was produced. 
In this all-liquid TPC configuration, the high voltage system developed similar instabilities at a few kV higher voltages on the cathode than with a gas electrolumiscence region. 
Therefore, we attribute this phenomenon to the high voltage region in the argon-xenon mixture liquid. 
This instability limits the detector operation to relatively high pressures and will continue to be investigated in future studies.  

\section{Conclusion}
\label{sec:concl}

We deployed a dual-phase TPC in a xenon-doped argon environment and studied the gain and pulse shape of gas electroluminescence signals excited by gamma radiation at different xenon concentrations. We observed a substantial increase (by $\sim$2.5) in the detected photon signal strength by VUV-sensitive SiPMs. In particular, data from a SiPM covered by a quartz window confirms significant Xe$^*_2$ light emission in the TPC starting at 1\% [Xe], suggesting efficient energy transfer from argon excitation to xenon in the electroluminescence process. The energy transfer and light emission mechanism is quantitatively studied with an analytical model. The model successfully reproduces the evolving electroluminescence waveform observed by SiPMs with and without quartz windows. % and the relative signal distribution between different SiPM units.
In addition, we also report that xenon-doped liquid argon at the percent level is a potent wavelength shifter that can strongly absorb short wavelength photons and emit them at longer wavelength. This observation enables the use of conventional reflectors to improve the electroluminescence energy resolution for liquid argon. The leakage of this wavelength-shifted light into the gaseous region is found to be relative small, allowing accurate position reconstruction with direct photon signals. %\remark{Jimmy: Some of the above discussion is appendix exclusive. do we modify/omit some of these statements to reflect that? I think not but want to bring it to attention.}

\begin{acknowledgments}

 This project is partially supported by the U.S. Department of Energy (DOE) Office of Science, 
 Office of High Energy Physics under Work Proposal Number SCW1676 and SCW1504 awarded to Lawrence Livermore National Laboratory (LLNL). 
JWK and JQ have been supported by the HEPCAT Consortium awarded to the University of California under award no DE-SC0022313. JQ was also supported by the Office of Science Graduate Student Research Fellowship by the US DOE. 

We thanks Sergey Pereverzev for discussions on topics related to this work. 

This work was performed under the auspices of the U.S. Department of Energy by Lawrence Livermore
National Laboratory under Contract DE-AC52-07NA27344. 
LLNL release number: LLNL-JRNL-2010955. 

\end{acknowledgments}

\appendix

\section{S2 Waveform Modeling Details}
\label{appendix:avg_wf_appendix}

\subsection{Event Selection}

\begin{figure}[!htbp]
    \centering
    \includegraphics[width=0.9\linewidth]{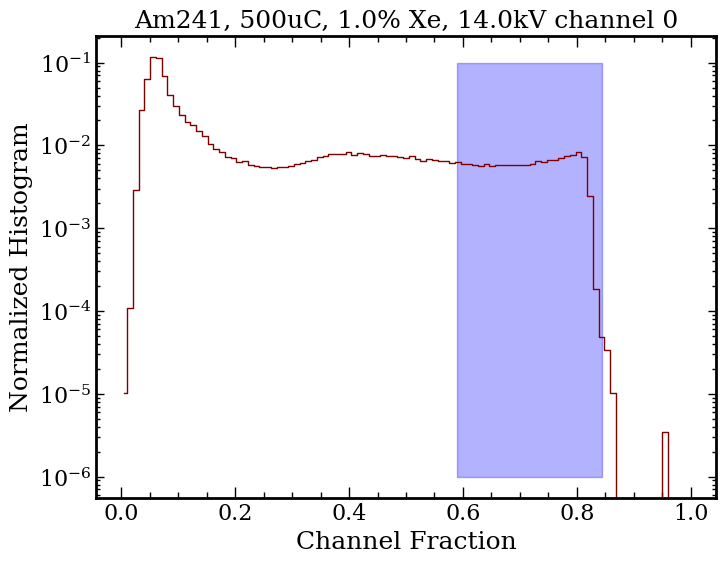}
    \includegraphics[width=0.9\linewidth]{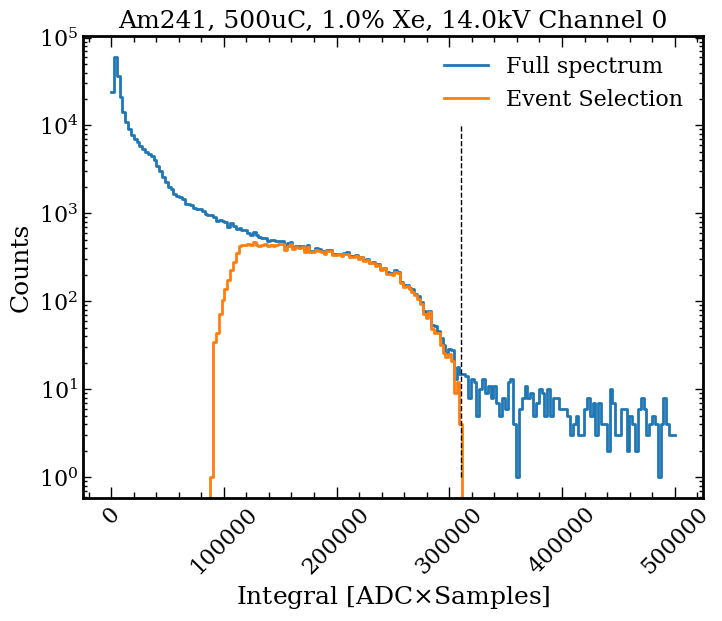}
    \caption{\textbf{Top}: Distribution of $f_0$ for data taken at 14~kV on the cathode with a 500~$\mu$Ci $^{241}$Am source. The shaded blue region indicates where $f_0\in[0.7f_{0,max}, f_{0,max}]$, but the actual selection is where $f_0>0.7f_{0,max}$. \textbf{Bottom}: The full event selection, as described in the text, used to compute the average waveform for Ch 0.}
    \label{fig:fraction_cuts}
\end{figure}

The average waveforms for top SiPMs are ideally computed for EL events occurring right below the SiPM under investigation. 
In our simple TPC, such events are selected with 
%In computing the average waveform, we align the individual event waveforms by the timestamp at which the pulse reaches 50\% of its maximum amplitude, known as $t_{half}$. To obtain an accurate measure of $t_{half}$, we need to select large pulses with sufficient numbers of photons. For each top channel, 
the fraction of S2 light seen by the top SiPM channel in question, $i$, 
\begin{equation}
    f_i = S2_i/S2_{total}
\end{equation} 
Fig.~\ref{fig:fraction_cuts} (top) shows an example $f_i$ distribution. 
In this analysis we select events with $f_i>0.7f_{i,max}$, where $f_{i, max}$ is the maximum $f_i$ value for the distribution. 
Because $f_{i, max}$ varies for different SiPM channels, the minimum $f_i$ value ranges between 0.45 for Ch 1 and 0.59 for Ch 0. 
The effect of this cut can be seen in the bottom subfigure, where only large area pulses are selected. 
The use of large pulses with sufficient numbers of photons also ensures an accurate calculation of $t_{half}$ -- the time a pulse reaches 50\% of its maximum amplitude, which is used to align individual event waveforms for averaging. 
We also reject events with sizes greater than the $^{241}$Am end-point (Fig.~\ref{fig:exFits}), which are attributed to background events with less understood origins.

%When we look at the integral spectrum for each top channel pulse, we notice that $^{241}$Am events have a clear drop-off in counts past a certain integral value (Fig.~\ref{fig:exFits}). Events with an integral greater than the $^{241}$Am end-point are background events, and a maximum integral cut is also applied to each of the top channels. We know this since we took data with both a 10~$\mu$Ci and a 500~$\mu$Ci source placed at the same location, and the end-point was far clearer with the 500~$\mu$Ci source.

\begin{figure}[!t]
    \centering
    \includegraphics[width=0.9\linewidth]{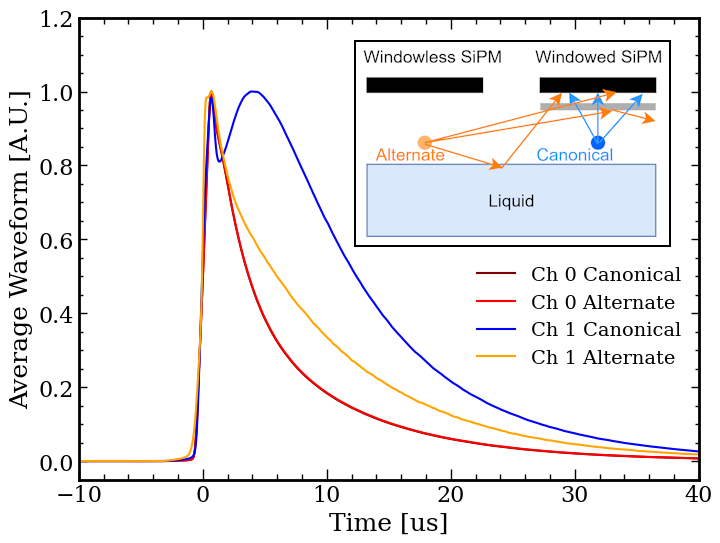}
    \caption{\textbf{Inset}: An illustration of the difference between canonical and alternate events. For the windowed SiPM, canonical events occur beneath the windowed SiPM itself, and the photons from the S2 signal have a relatively direct path towards the SiPM. Alternate events are those that are under either of the windowless SiPMs. As such, the light that reaches the windowed SiPM either strike at a glancing angle, or come from wavelength shifting at the liquid-gas interface. \textbf{Main figure}: The average waveforms for the top windowless (Ch 0) and windowed (Ch 1) SiPMs computed with the alternate and canonical cuts for 1\% [Xe]. The dark red and bright red lines are practically identical. We explain the reason for the difference between the average waveforms in the windowed SiPM using these two  event selections in the text.}
    \label{fig:canonical_alternate_avgwf}
\end{figure}

Following the same principle, the average waveforms for the bottom channels are computed using the same events selected as for the channel directly above them in the gas~\footnote{The top channels, 0, 1, and 2, are directly above the bottom channels, 4, 3, and 5, respectively.}. However, these ``canonical'' selections do not work for Chs 1 and 3 in pure argon because the quartz window of Ch 1 blocks argon EL light and does not collect significant S2 light. Instead, for pure argon, the events used to compute the average waveforms for Chs 1 and 3 are those for which \textit{either} Ch 0 or Ch 2 has $f_i>0.7f_{i,max}$; we refer to this cut as the ``alternative'' selection. 
%\subsection{An Alternative Event Selection}
%To summarize the previous section, the top windowed SiPM's average S2 waveforms are computed using events under this SiPM (``canonical'' selection) for data with xenon doping, but using events below neighboring windowless SiPMs (``alternative'' selection) for pure argon data. 
The two event selection criteria are illustrated in the inset of  Fig.~\ref{fig:canonical_alternate_avgwf}, and are used for the main analysis presented in Sec.~\ref{sec:s2_shape}. 

We also studied the alternative S2 waveforms at 1\% [Xe] for top SiPMs Chs 0 and 1, which are compared to the canonical ones in Fig.~\ref{fig:canonical_alternate_avgwf}. 
With the alternative event selection, the S2 pulses are still produced in the central region of the detector with well defined gain properties, as confirmed by the consistent waveforms for the top windowless SiPM (Ch 0) calculated in these two different ways. 
Direct Xe$^*_2$ S2 light emitted from a distance, however, would only hit the quartz window of Ch 1 SiPM at large incidence angles, with a higher probability of reflecting off of the quartz window, and is thus suppressed. 
As a result, the double-hump structure in the canonical windowed SiPM waveform is missing from the alternative windowed SiPM waveform. 
Instead, we observe a shape similar to that of the windowless SiPMs, suggesting that some Ar$_2^{*3}\Sigma$ (the triplet state) light may be wavelength-shifted in the liquid mixture and then arrive at the windowed SiPM window in a more favorable, non-glancing angle. 
It is noteworthy that the alternative Ch 1 waveform at 1\% [Xe] is nearly identical to those of the bottom SiPMs (Fig.~\ref{fig:all_wfs_1p_xe}), which are dominated by wavelength-shifted Ar$_2^*$ light.

\subsection{Model Fit and Uncertainties}

\begin{figure*}[!htb]
    \centering
    \includegraphics[width=0.8\linewidth]{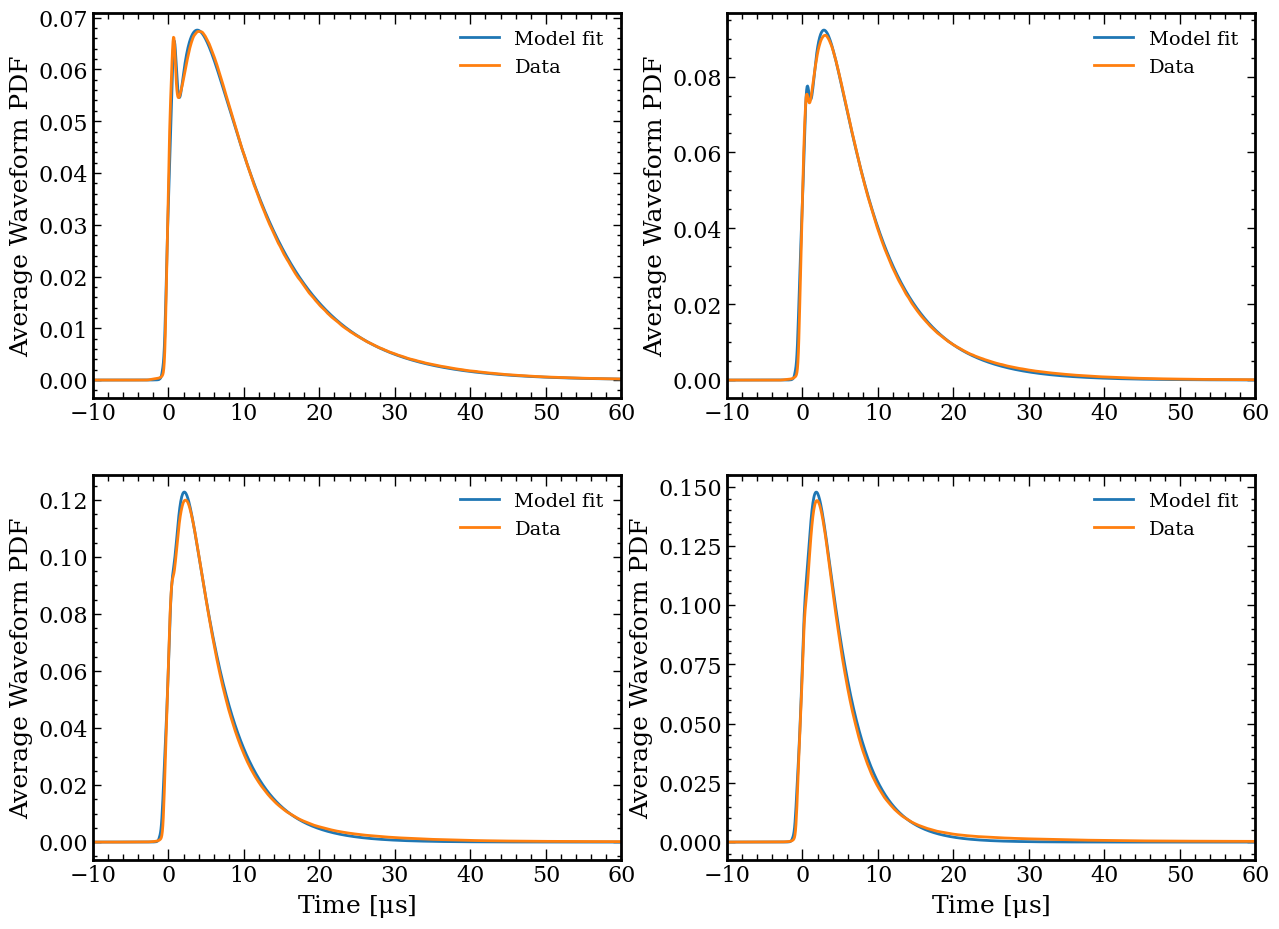}
    \caption{Fits of the EL model to the top windowed SiPM (Ch 1) across all doping concentrations from 1\% to 4\%. The subplots are arranged in increasing [Xe] from left to right and up to down. We see that the hump structure is well modeled as the difference of two exponentials.}
    \label{fig:windowed_sipm_avgwf_fits}
\end{figure*}

Instead of separately modeling the longitudinal electron cloud diffusion in the liquid and the electron transit through gas described in  Eq.~\ref{eq:full_s2_pulse}, we conveniently use the average neutral bremsstrahlung waveform observed by the top windowed SiPM in pure argon as a template for $f_{kernel}(t) = f_{SPE}(t)*f_{SE}(t)*\text{Gauss}(t|z)$ (averaged across different values of $z$). %The systematic uncertainties of this approach are discussed later in this section.
$P_{light}(t)$ is treated as a linear combination of the neutral bremsstrahlung (modeled as a delta function), Ar$_2^*$ de-excitation (Eq.~\ref{eq:ar_timing}) and the Xe$^*_2$ time profile (Eq.~\ref{eq:xe_timing}). 
Then $f_{kernel}(t)$ is convolved with $P_{light}(t)$ to fit the average S2 waveforms in xenon-doped argon. 

In the handling of model parameters, because $k_1$, $k_2$, $\tau_2$ and $\tau_3$ only appear in the combinations: $1/\tau_2 + k_1n_{Xe}\equiv r_A$ and $1/\tau_3 + k_2n_{Xe}\equiv r_B$, we fit $r_A$ and $r_B$ for each liquid xenon concentration. %$k_1$ ($k_2$) and $1/\tau_2$ ($1/\tau_3$) can then be inferred from the slope and intercept of $r_A$ ($r_B$) as a function of $n_{Xe}$, as illustrated in Fig.~\ref{fig:k1_k2_linearity} (bottom). 
In total, there are four free parameters in the fit for each average waveform in the top windowed SiPM: $r_A$, $r_B$, the Xe$_2^*$ excimer amplitude, and an Ar$^*_2$ excimer amplitude corresponding to the liquid wavelength-shifting effect; since the average waveform area is normalized to 1, a parameter for the neutral bremsstrahulung amplitude would be redundant. 
Figure~\ref{fig:windowed_sipm_avgwf_fits}  shows the fit results for the average waveforms for the top windowed SiPM Ch 1 across all xenon concentrations using the model. 

The best-fit $r_A$, $r_B$ values are shown in Fig.~\ref{fig:k1_k2_linearity} (bottom) as a function of xenon concentration in the liquid. The observed linearity of these values with xenon concentration confirms that the rate of energy transfer from Ar$_2^*$ to Xe and the rate of formation of Xe$_2^*$ from atomic Xe excitation are both proportional to the concentration of xenon. 
The gaseous xenon concentration values are directly measured in the experiment with a calibrated RGA system, yielding an average $14\pm4$~ppm increase in the gas for every 1\% [Xe] in the liquid. 
We estimate $k_1 = (1.1\pm0.3)\times 10^{-10}$~cm$^3$/s and $k_2 = (2.8\pm0.9)\times 10^{-11}$~cm$^3$/s. It should be noted that the 4~ppm uncertainty in the gaseous xenon concentration is a shared instrument uncertainty observed across many different [Xe] values. As such, the errors reported for $k_1$ and $k_2$ are conservative. 
The linear function intercepts at 0\% [Xe] also predict an Ar$_2^*$ triplet half life of $3.2\pm0.8~\mu\text{s}$ and a Xe$^*$ (or ArXe$^*$) de-excitation time of $26\pm8~\mu\text{s}$ at liquid argon temperature. 

\begin{figure*}[!htb]
    \centering
    \includegraphics[width=0.9\linewidth]{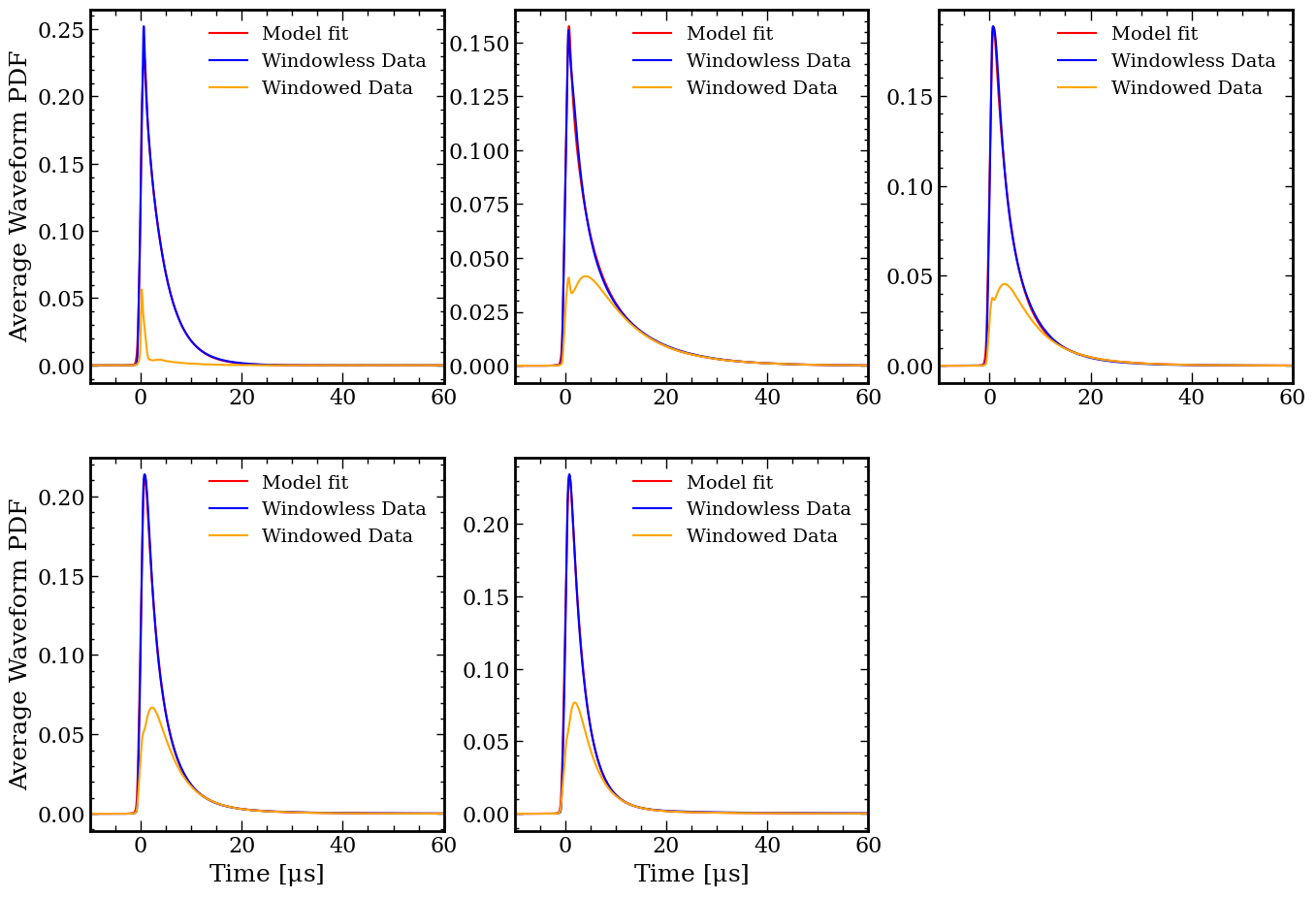}
    \caption{Fits of the EL model to the top windowless SiPM (Ch 0). The plots are arranged in increasing [Xe] from left to right, top to bottom, where the top left plot is for 0\% [Xe] in liquid, and the bottom right plot is for 4\% [Xe] in liquid. We can see the model fits the data reasonably well, and that the tails of the windowed and windowless SiPMs match.}
    \label{fig:windowless_sipm_avgwf_fits}
\end{figure*}

The top windowless SiPM's average S2 waveform is modeled similarly to the top windowed SiPM waveform, but with much enhanced contributions from Ar$^*_2$ luminescence (Eq.~\ref{eq:ar_timing}) because of its direct sensitivity to 128~nm photons. 
The windowless SiPMs can also detect additional 147~nm light but its time profile should be proportional to that of Xe$_2^*$ (Eqs.~\ref{eq:xe_timing} and \ref{eq:147nm_timing}).
We use the top windowed SiPM waveform as a template in the fit to absorb contributions from both 147~nm and 175~nm light.
This fit contains three free parameters: $r_A$ as described above, the windowed SiPM waveform amplitude, and the amplitude of the singlet Ar$^*_2$ component (approximated as a delta function); the amplitude for the Ar$^*_2$ triplet is not necessary due to the normalized waveform area.  
The fit results are shown in Fig.~\ref{fig:windowless_sipm_avgwf_fits}. %We comment that the $r_A$ value extracted from the windowless SiPM fit does not directly measure the Xe$_2^*$ formation process,
%and is not used for inferring the physical quantities such as $k_1$. 

%\subsection{Unmodeled Physics}

\begin{figure*}[!htbp]
    \centering
    \includegraphics[width=0.4\linewidth]{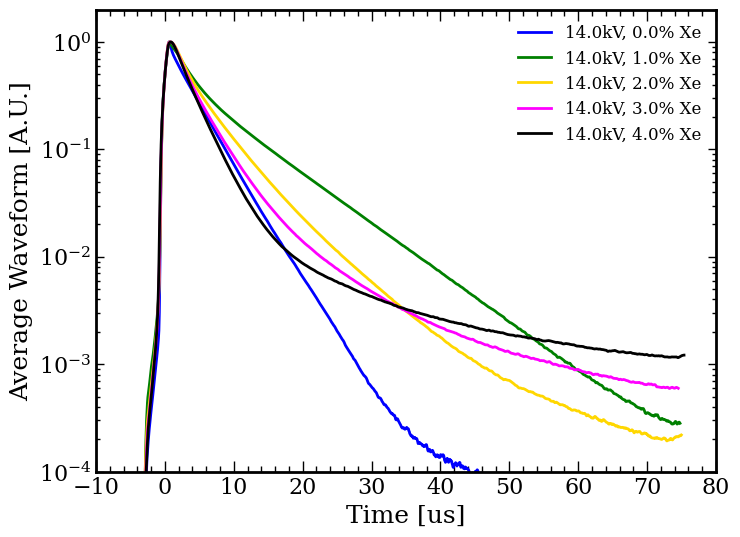}
    \includegraphics[width=0.4\linewidth]{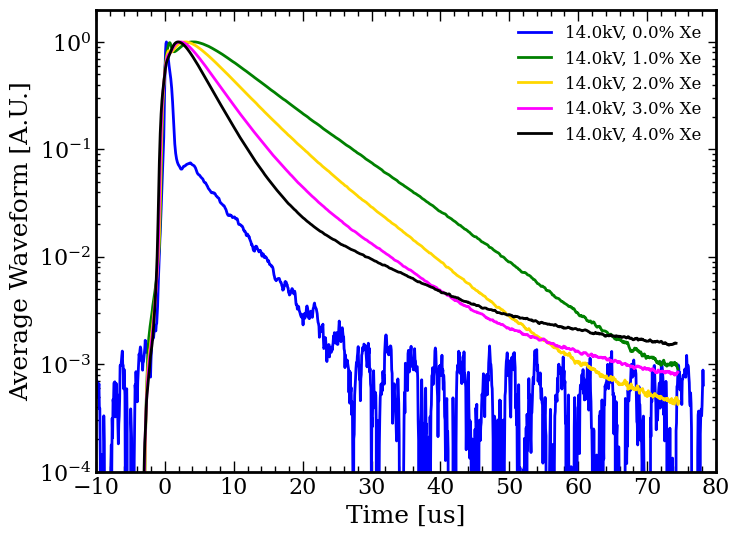}
    \caption{The top windowless (left) and top windowed (right) SiPM waveforms plotted in log scale so that the long tails are made apparent.
    }
    \label{fig:average_waveforms_14kv_tails}
\end{figure*}

%The model presented in Sec.~\ref{sec:modeling_s2} only describes the effects that have strong observational evidence and an identifiable physical origin. These include: the two-step transfer of energy from Ar$_2^{*3}\Sigma$ to Xe$^*$ to Xe$_2^*$; the wavelength shifting of Ar$_2^*$ light into Xe$_2^*$ light by the high [Xe] in the liquid; and the presence of neutral bremsstrahlung light seen by the windowed SiPMs. H
Despite its success in describing leading order S2 pulse shape observations, the model proposed in Fig.~\ref{fig:arxe_scintillation_gas} is incomplete. 
For example, a long, slowly decaying tail  is observed in the average S2 waveforms of both the windowed \textit{and} windowless SiPMs when xenon is added, as illustrated in Fig.~\ref{fig:average_waveforms_14kv_tails}. Prior studies on scintillation from gaseous argon-xenon mixtures have observed and/or proposed extra de-excitation pathways, some of which are summarized in Fig.~18 of \cite{HBrunet_1982}. 
Such energy transfers are not considered in our model and their omission can cause the fit results to be inaccurate. In fact, a different Ar$_2^{*3}\Sigma$ to Xe$^*$ energy transfer rate ($k_1$) is obtained using the windowless SiPM fit from that using the windowed SiPM fit. 
Because the windowless SiPM accepts VUV photons below 160~nm wavelength, its collected signals are more likely to include unmodeled higher-orders effects. In addition, the absence of the slow hump in the windowless SiPM waveforms means that the fitted Ar$_2^{*3}\Sigma$ to Xe$^*$ energy transfer rate is inferred from the Ar$_2^{*3}\Sigma$ decay rate, which could differ from the Xe$^*$ buildup rate because of contributions from additional Ar$_2^{*3}\Sigma$ de-excitation pathways. 
%while the windowed SiPM is primarily sensitive to de-excitation pathways that yield Xe$_2^*$, which is modeled in our work.

%However, introducing all these pathways into our model would cause the fits to be under-constrained. 

Adding extra pathways to our model would likely \textit{improve} the fit to our data, and may explain the long tails in Fig.~\ref{fig:average_waveforms_14kv_tails}. 
However, this approach could cause the fits to be under-constrained due to our limited wavelength selectivity. 
More importantly, the physical origin of some of the extra de-excitation pathways proposed in \cite{HBrunet_1982} is in slight tension with other studies. For example, \cite{10.1063/1.434079} did not report an observation of 129.6~nm light from the direct de-excitation of Xe($^1P_1$), while \cite{HBrunet_1982} did. 
Therefore, we consider such extensions to our model beyond the scope of this work. 

\begin{figure}[!htbp]
    \centering
    \includegraphics[width=0.9\linewidth]{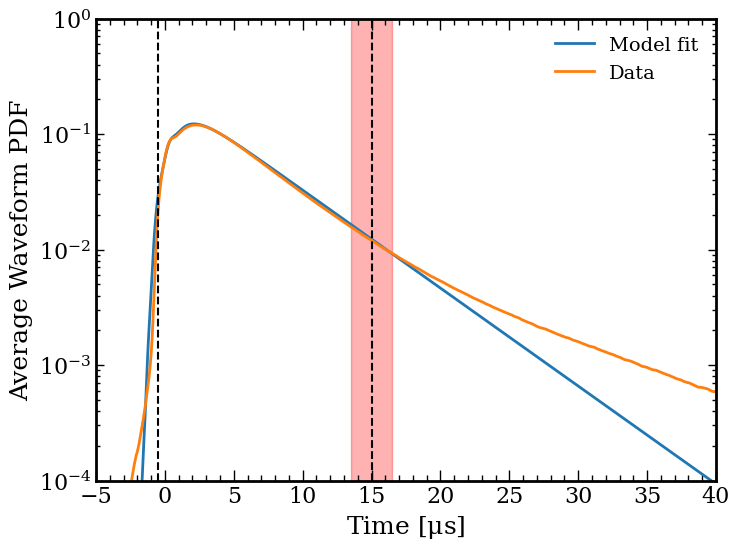}
    \caption{The top windowed SiPM's average S2 waveform and its fit at 3\% [Xe] in the liquid. The black dashed lines show where the waveform fit is performed, and the red band shows the range at which the cut-off is varied.}
    \label{fig:cutoff_fig}
\end{figure}

There are two main sources of systematic uncertainty in fitting the average waveform: the choice of time range used in the fit, and the choice of the kernel used to convolve $P_{light}(t)$ to get the full average waveform model. The time range used for the fit starts at -0.5~$\mu$s (-0.25~$\mu$s for the 2\% [Xe] data) and is cut off right around when the waveform reaches the unmodeled long tail (Fig.~\ref{fig:cutoff_fig}). This cut-off value is 35~$\mu$s, 20~$\mu$s, 15~$\mu$s, and 12~$\mu$s for the average waveforms corresponding to 1-4\% [Xe] in the liquid, respectively. As these cut-offs are somewhat arbitrary, we perform the fit for five values between $\pm10\%$, inclusive, of the nominal cut-off value, as shown in Fig.~\ref{fig:cutoff_fig}.

\begin{figure}[!htbp]
    \centering
    \includegraphics[width=0.9\linewidth]{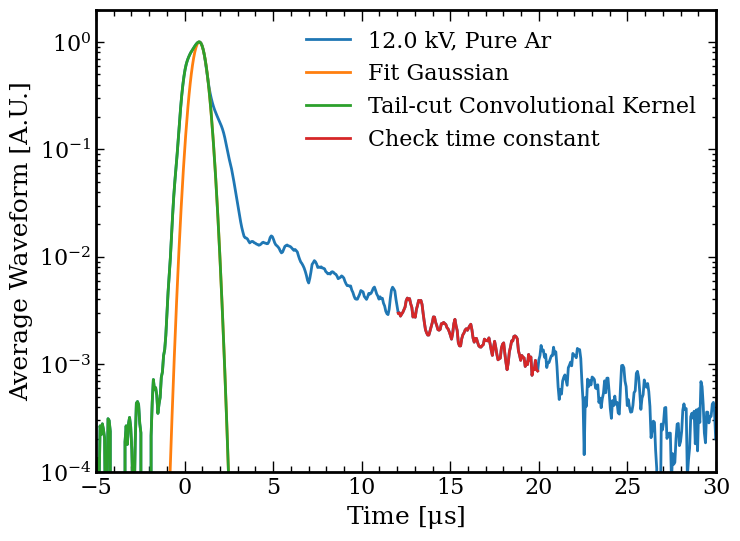}
    \caption{The $f_{0tw}(t)$, as defined in the text, is shown in blue. $f_{kernel}(t)$ can be defined by either cutting off $f_{0tw}$ at around 27~$\mu$s, where it returns to baseline, or by fitting the falling edge after the maximum of $f_{0tw}$ with a gaussian, and stitching it to the rising edge of $f_{0tw}$, as shown in green.}
    \label{fig:cutoff_kernel_fig}
\end{figure}

For the convolution kernel, $f_{kernel}(t)$, we actually use the top windowed SiPM's pure argon average waveform\footnote{The subscript ``0tw" means ``0\% Xe top windowed".}, $f_{0tw}(t)$, at 12~kV on the cathode instead of at 14~kV, since the 14~kV average waveform features a prominent tail as shown in Fig.~\ref{fig:neutral_brem_hypothesis}. 
Upon plotting in log-scale, $f_{0tw}(t)$ at 12~kV \textit{also} features a tail, albeit substantially smaller than at 14~kV. Other groups have reported that SiPM afterpulsing contains one or more $\mu$s-scale exponential components \cite{X-ArT:2024npd}, which may explain our observed tail. We use two methods to compute $f_{kernel}(t)$ to address this uncertainty. The first way involves fitting the portion of $f_{0tw}(t)$ immediately after its maximum with a gaussian, effectively cutting off the long tail. The second way keeps the long tail, and cuts off $f_{0tw}(t)$ close to when the waveform returns to baseline. Both choices are shown in Fig.~\ref{fig:cutoff_kernel_fig}. To obtain mean and uncertainty of the fit parameters at each [Xe] value, we take the average and standard deviation of all ten configurations corresponding to the five choices of fit-range cut-off values and two choices of $f_{kernel}(t)$.

\subsection{Evidence for Wavelength Shifting}

Sec.~\ref{sec:modeling_s2} primarily focuses on the \textit{shape} of our S2 waveforms. 
However, our model can also estimate the amounts of  Ar$_2^*$ light, 147~nm light, and Xe$_2^*$ light emitted from the gas gap by integrating Eqs.~\ref{eq:ar_timing}, ~\ref{eq:xe_timing}, and ~\ref{eq:147nm_timing}. 
\begin{subequations}
    \begin{equation}
        I_{Ar} = N_{Ar} \left(p_S + (1-p_S)\frac{1}{1+k_1\tau_2n_{Xe}}\right)
    \end{equation}
    
    \begin{equation}
        I_{147} = N_{Ar}(1-p_S) \frac{1}{1+\frac{1}{\tau_2 k_1 n_{Xe}}}\frac{1}{1+\tau_3 k_2 n_{Xe}}
    \end{equation}
    
    \begin{equation}
        I_{Xe} = N_{Ar}(1-p_S) \frac{1}{1+\frac{1}{\tau_2 k_1 n_{Xe}}}\frac{1}{1+\frac{1}{\tau_3 k_2 n_{Xe}}}
    \end{equation}
    \label{eqs:integrated_ratios}
\end{subequations} 
These equations predict that the amount of Ar$_2^*$ light monotonically decreases with $n_{Xe}$, the 147~nm light first increases then decreases with $n_{Xe}$, and the Xe$_2^*$ light monotonically increases with $n_{Xe}$. 
In principle, the ratios between these integrated light intensity predictions could be compared to the ratio of the light seen by the top windowed SiPM to the light seen by the top windowless SiPM, $r_{w/wl}$, at each [Xe] level. 
This can, \textit{naively}, be calculated as \begin{equation}
    r_{w/wl} = \frac{T_{Xe}q_{Xe}I_{Xe}}{q_{Xe}I_{Xe}+q_{147}I_{147}+q_{Ar}I_{Ar}}
    \label{eq:r_w_wl_naive}
\end{equation} where $q_i$ is the quantum efficiency of each SiPM to component $i$, and $T_{Xe}$ is the transmission of the quartz window to Xe light. 

This calculation does not consider the wavelength-shifted Ar$_2^*$ light detected by the windowed SiPM, which is not included in our energy transfer model. 
As a result, it \textit{under-predicts} $r_{w/wl}$, especially at low [Xe] values.  
For example, Eq.~\ref{eq:r_w_wl_naive} predicts $r_{w/wl}\approx0.3$ for 1\% [Xe], while the $^{241}$Am endpoint analysis in Sec.~\ref{sec:electroluminescencegainResults} yields a value of 0.55.
However, wavelength-shifted Ar$_2^*$ light is detected by the windowed SiPM in our experiment, as indicated by studies of the top windowed SiPM waveform with distant S2s (Fig.~\ref{fig:canonical_alternate_avgwf}), and of the bottom windowed SiPM signals (Fig.~\ref{fig:all_wfs_1p_xe}). 
Quantitatively, the windowed SiPM waveform fits include this component with a floating amplitude, resulting an estimate of 7--20\%.

We also studied the wavelength shifting phenomenon in the liquid mixture with a simplified Geant4-based optical Monte Carlo simulation. 
The simulation assumes an S2 wavelength composition based on the model predictions described in Eqs.~\ref{eqs:integrated_ratios}, with the model parameters fixed at their best-fit values. 
Then, all 128~nm and 147~nm light entering the liquid is absorbed and remitted isotropically, and the wavelength-shifted photons are tracked until their detection by SiPMs or absorption by detector materials. 
This simulation also estimates that $\sim$20\% of the photons detected by the top windowed SiPM can be from the liquid wavelength shifting process. 
It raises the predicted $r_{w/wl}$ value to $\sim$0.4, closer to that measured in Sec.~\ref{sec:electroluminescencegainResults}.  

\begin{figure}[!htb]
%    \centering
%    \includegraphics[height=0.22\textheight]{sim_vs_data_pure_ar_aft.png}
%    \includegraphics[height=0.22\textheight]{aft_dist_1p_xe_sim_vs_data.png}
    \includegraphics[width=0.9\linewidth]{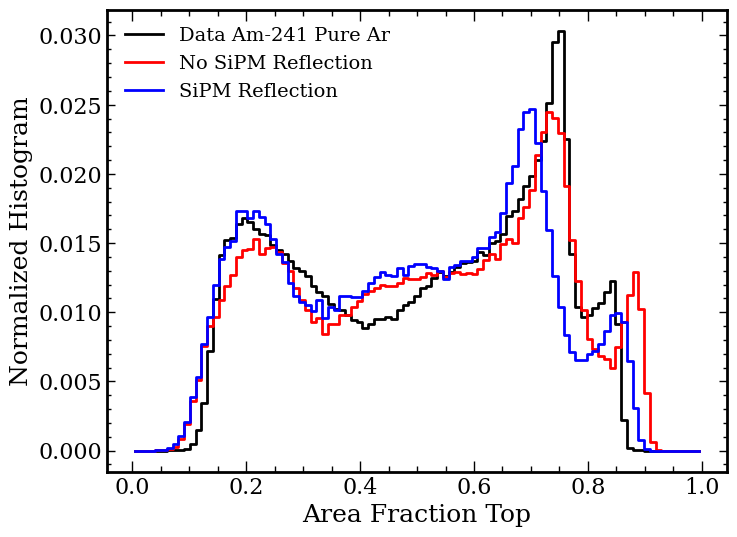}
    \includegraphics[width=0.9\linewidth]{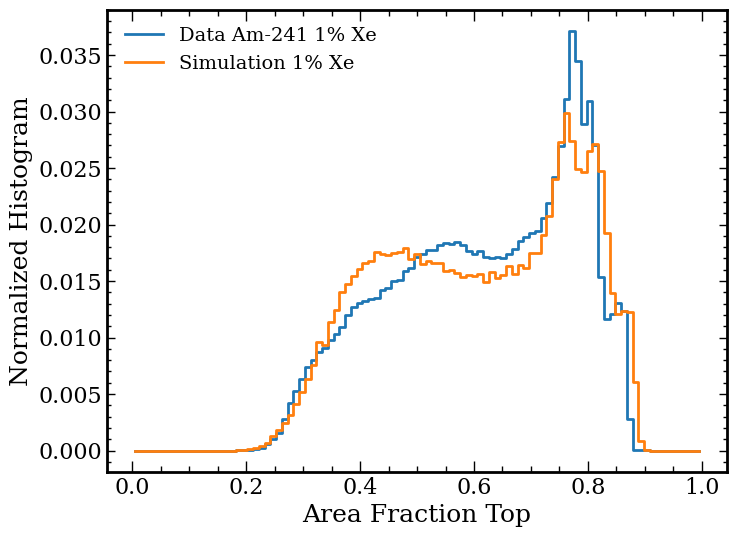}
    \caption{\textbf{Top:} A comparison of the simulated AFT distributions with data (black histogram) for pure argon. The red simulation result ignores reflection of 128~nm light on SiPMs, while the blue one assumes the same reflectivity as measured for 175~nm light in liquid xenon~\cite{nEXO:2021uxc} (no 128~nm measurement exists to our knowledge). \textbf{Bottom:} Simulated AFT distribution with wavelength shifting of 128~nm and 147~nm light and SiPM reflection in liquid argon with 1\% [Xe].}
    \label{fig:sim_data_aft_comparison}
\end{figure}

In addition, the optical simulation also produces a distribution for the S2 light split between the top and bottom SiPM assemblies. 
Fig.~\ref{fig:sim_data_aft_comparison} shows the distributions of the predicted fraction of S2 light seen by the top SiPM assembly (Area Fraction Top or ``AFT'') for pure argon (no wavelength shifting) and for 1\% [Xe] (with wavelength shifting), compared to those measured in the data. 
The simulation reproduces the main features of the data distributions. 
For pure argon, the main peak at AFT$\sim$0.7 corresponds to S2s produced below and around the two top windowless SiPMs (Ch 0 and Ch 2), with events near the center of SiPM Ch 0 giving another higher AFT peak due to its higher light collection efficiency from the compactness of the SiPM sensitive area. The low AFT peak around 0.2 is from events below and around the windowed SiPM Ch 1, which is opaque to Ar$^*_2$ light. 
Tuning the SiPM reflectivity for VUV photons around the reflectance values measured by the nEXO collaboration \cite{Lv:2019res,nEXO:2021uxc} shifts the AFT distribution but does not produce a perfect match. 
For the 1\% [Xe] data, the windowed SiPM detects significant light, eliminating the AFT peak around 0.2. 

The simulation-data agreement may be further improved by incorporating higher-order optical physics, such as the angular-dependent reflectance of VUV light on the SiPM surfaces in gas and in the liquid mixture. However, data on such effects are not available. 
In addition, the relative amounts of 128~nm, 147~nm, and 175~nm light, which are inputs for our optical simulation, are estimated from our simplified EL model; these inputs are also subject to uncertainties from the omission of additional energy transfer and de-excitation pathways discussed in the previous section. 

\newpage
\bibliographystyle{apsrev}
\bibliography{biblio}

\end{document}